\documentclass[usenatbib]{mnras}
\usepackage[T1]{fontenc}
\usepackage[utf8]{inputenc}
\usepackage{blindtext}
\usepackage{comment}
\usepackage{amsmath}
\usepackage[dvipsnames]{xcolor}
\usepackage{array}
\usepackage{graphicx}
\usepackage{tikz}
\newcolumntype{P}[1]{>{\centering\arraybackslash}p{#1}}

\title[Unequal-mass MBHB signatures]{Thermal X-ray signatures in late-stage unequal-mass massive black hole binary mergers}

\author[Krauth et al.]{ Luke Major Krauth$^{1,2}$\thanks{E-mail: LMK2202@columbia.edu},
Jordy Davelaar$^{3,4}$, 
Zoltán Haiman$^{2,5,6}$,
John Ryan Westernacher-Schneider$^{7}$, \and
Jonathan Zrake$^{8}$
and Andrew MacFadyen$^{9}$
\\
$^{1}$Gravitation Astroparticle Physics Amsterdam (GRAPPA), University of Amsterdam,\\
Science Park 904, 1098 XH Amsterdam, The Netherlands\\
$^{2}$Department of Physics, Columbia University, New York, NY 10027, USA\\
$^3$Department of Astrophysical Sciences, Peyton Hall, Princeton University, Princeton, NJ 08544, USA\\
$^{4}$NASA Hubble Fellowship Program, Einstein Fellow\\
$^{5}$Department of Astronomy, Columbia University, New York, NY 10027, USA\\
$^{6}$Institute of Science and Technology Austria (ISTA), Am Campus 1, Klosterneuburg 3400, Austria\\
$^{7}$Westgate Research, PO Box 181, Cardiff, ON K0L 1M0, Canada\\
$^{8}$Department of Physics and Astronomy, Clemson University, Clemson, SC 29634, USA\\
$^{9}$Center for Cosmology and Particle Physics, Physics Department, New York University, New York, NY 10003, USA\\
}

\date{Accepted XXX. Received YYY; in original form ZZZ}
\pubyear{2025}

\begin{document}
\maketitle

\begin{abstract}
\normalsize
The multi-messenger combination of gravitational waves (GWs) from merging massive black hole binaries (MBHBs) and the electromagnetic (EM) counterpart from the surrounding circumbinary disk (CBD) will open avenues to new scientific pursuits. In order to realize this science, we need to correctly localize the host galaxy of the merging MBHB. Multi-wavelength, time-dependent electromagnetic (EM) signatures can greatly facilitate the identification of the unique EM counterpart among many sources in LISA's localization volume. To this end, we studied merging unequal-mass MBHBs embedded in a CBD using high-resolution 2D simulations, with a $\Gamma$-law equation of state, incorporating viscous heating, shock heating and radiative cooling. We simulate each binary starting from before it decouples from the CBD until just after the merger. We compute EM signatures and identify distinct features before, during, and after the merger. We corroborate previous findings of a several order of magnitude drop in the thermal X-ray luminosity near the time of merger, but with delayed timing compared to an equal-mass system. The source remains X-ray dark for hours post-merger. Our main results are a potential new signature of a sharp spike in the thermal X-ray emission just before the tell-tale steep drop occurs. This feature may further help to identify EM counterparts of LISA's unequal MBHBs before merger without the need for extensive pre-merger monitoring. Additionally, we find a role-reversal, in which the primary out-accretes the secondary during late inspiral, which may diminish signatures originating from Doppler modulation.

\end{abstract}


\section{Introduction}

Stemming from the fundamental work of quasar observations in the 1960s \citep{Schmidt1963}, astronomers theorized that most galaxies likely host a massive black hole (MBH) at their cores \citep{LyndenBell1969}. 
On the other hand, galaxies merge regularly over cosmic time. Thus, it is expected that the MBHs hosted by two merged galaxies eventually find each other and form a massive black hole binary (MBHB) \citep{Begelman1980}.
For MBHs, gas surrounding the central mass is expected to radiate efficiently and collapse along its rotation axis, supported by angular momentum, and therefore turn into a thin disk \citep{Prendergast1968,Pringle1972,Shakura1973}. A similar outcome is expected for the gas that surrounds a MBHB after a galaxy merger, with the gas forming into what is known as a circumbinary disk (CBD).

In the effort to detect these MBHBs, the Laser Interferometer Space Antenna (LISA; \citealt{Amaro-Seoane2023,Colpi2024}) will be launched next decade. LISA is aimed to measure gravitational waves in the range of $\sim$0.1~mHz and $\sim$1~Hz, residing in the current gap in detectable frequencies by the LIGO-Virgo-KAGRA collaboration \citep{Aasi2015,Acernese2015,Aso2013}
and that of the International Pulsar Timing Array collaboration \cite[IPTA;][]{IPTA2023}. LISA will be sensitive to gravitational waves (GWs) from MBHBs with total masses of $\sim10^3-10^7 M_{\odot}$, probing the range of MBHBs expected from galaxy mergers \citep{Amaro-Seoane2023,Colpi2024}.

Gravitational wave observations alone can provide detailed information about the component masses, orbital parameters, spin characteristics, and luminosity distance of the MBHB \citep{Bogdanovic2022}. However, when combined with the information obtained from the surrounding CBD via electromagnetic (EM) signatures, such as the spectral energy distribution (SED), periodic variability, jet orientation, and also orbital parameters, significantly more can be achieved. By timing the gravitational wave (GW) signal relative to the EM signal one can constrain the speed of gravitational waves \citep{Haiman2009}, thereby testing General Relativity against alternative theories, such as those requiring extra dimensions \citep{Rham2018}, or massive gravity \citep{Hassan2012}. Further, EM signals can allow for the unique identification of the host galaxy in LISA's error volume. Concrete identification will provide a relation between merging MBHs and their host galaxies as a function of redshift, luminosity and other properties, allowing us to better understand the co-evolution of MBHs with their host galaxies \citep{Kormendy2013}, as well as gain insight into the primary mechanisms driving the mass growth of SMBHs over cosmic epochs (see, e.g., \citealt{Baker2019,Bogdanovic2022} for reviews and references). 
As such, understanding the potential EM signatures originating from the circumbinary gas is paramount for these scientific goals.

In recent years, numerical simulations have converged on some aspects of these MBHB+CBD systems. The orbiting MBHB carves out a surrounding, low density, elongated central cavity in the CBD \citep{Artymowicz1996,MacFadyen2008,Shi2012,Ragusa2016}. This cavity shape is self-reinforcing. As the binary orbits, it tidally strips material from the closest edge of the cavity wall, forming streams to feed the BHs. Some of this material encircles each BH and forms what are known as ``minidisks.'' Some of the stream is flung back out into the cavity, repetitively impinging on the cavity wall, most notably at the far edge of the cavity, contributing to the elongation of the cavity's shape. Additionally, a non-axisymmetric density buildup, commonly called the ``lump,'' can be produced. There are flow patterns in the cavity wall that are associated with the lump, which propagate around the binary once every several binary orbits \citep{Shi2012,Dorazio2013,Noble2021}. 
Any of these regions - streams, CBD, lump, minidisks - as well as their interplay - stream-stream interactions, stream-cavity interaction, lump migration, minidisks trading mass - can lead to observable signatures \citep{Westernacher-Schneider2022}. By modeling these systems and interactions, we can predict what those signatures should be.

A promising recent EM signature found by \cite{Krauth2023} is the disappearing thermal X-ray emission that occurs in many MBHB systems just before merger. Using high-resolution, 2D hydrodynamical simulations, \cite{Krauth2023} found that the minidisks surrounding each BH were responsible for the most energetic photon emissions in the system, namely falling in the X-ray band. As the binary inspirals due to GWs in late stages, the minidisks are tidally truncated. This leads to a reduction in their surface area, and, correspondingly, losses in their accretion rates and luminosity. As the system approaches merger, in the the last hours/days, this reduction is so great that there is a several-order of magnitude drop in the thermal X-ray luminosity. The accretion rate drop responsible for the X-rays' disappearance was also found in isothermal gas models \cite{Dittman2023}. \cite{Franchini2024} also confirmed this X-ray drop using 3D hyper-Lagrangian resolution simulations with up to 2.5th-order Post-Newtonian (PN) corrections. \cite{Clyburn2024}, studying more intermediate mass ratio inspirals (IMRIs) with mass ratios of $q=10^{-3}$ to $10^{-1}$ (where $q=M_2/M_1$ of the system, where $M_1$ and $M_2$ are the masses of the primary and secondary BHs, respectively) found similar thermal X-ray results in both light curves and spectra. Additionally, \cite{Zrake2025} confirmed this X-ray drop for certain models which depended on binary torque. Observing this distinct signature, which in principle only can be detected with as few as two data points (one before the drop and one after), could help LISA identify host galaxies of MBHB mergers without the need for extensive premerger monitoring that seeks to identify quasi-periodic oscillations. 

In their study, \cite{Krauth2023} varied the disk temperature, viscosity, and the total mass of the MBHB. However, one key parameter they did not vary was the mass ratio. This can be observationally important, as, while many MBHBs may be near equal mass, different $q$-values are to be expected. Cosmological simulations, such as the Illustris project \citep{Vogelsberger2014,Genel2014}, have shown that many ($\sim50\%$) of MBHB mergers should occur between $0.1<q<1$ \citep{Katz2020}. While prior studies have considered unequal mass ratios \citep{Clyburn2024,Zrake2025}, they have examined the IMRI regime and used an isothermal equation of state (EOS).
In this paper, we extend the work of \cite{Krauth2023} and others to include comparable mass cases with $q=0.1,0.3$ and $0.5$, including a more realistic gamma-law EOS. We use the same setup and code as \cite{Krauth2023}. Namely, we model the MBHB+CBD system with the 2D GPU-accelerated hydrodynamics code, \texttt{Sailfish}\footnote{https://github.com/clemson-cal/sailfish}. We incorporate physical viscosity, directly solve the energy equation, using a $\Gamma$--law equation of state for the gas, incorporate viscous heating, shock heating, and a physically-motivated cooling prescription. Late into inspiral the GW-driven inspiral time ($t_{\rm gw} \propto a^4$) decreases more rapidly than the viscous inflow time (e.g. $t_{\rm visc} \propto a^{7/5}$; \citealt{Pringle1991}). As such, it was initially thought that  the binary would outrun the circumbinary disk~\citep{Liu2003,Milosavljevic2005}, starving and dimming the binary. This limit is known as the ``decoupling limit.'' However, long-term 2D hydrodynamical simulations have shown that the circumbinary gas can follow the inspiraling binary and feed the individual BHs all the way until the merger~\citep{Farris2015a,Tang2018}. Nonetheless, we initialize inspiral from before the decoupling limit and run through post-merger. We calculate multi-wavelength light curves for the last hundreds of orbits through merger.

Our main new result is a potential new signature presenting in the lower range of unequal-mass binaries we study, in which there is a significant \emph{increase} in the thermal X-ray luminosity just before merger. This feature appears appreciably in our $q=0.3$ model and is even more prominent in our $q=0.1$ model. Regardless of pre-merger flare, there is the same characteristic, several-order-of-magnitude drop in the thermal X-ray luminosity when traversing from pre- to post-merger, previously reported for equal-mass mergers \citep{Krauth2023,Franchini2024}. However, the timing of this drop is delayed compared to equal-mass models. Additionally, we find a role-reversal occurs in late inspiral, in which the primary BH becomes the dominant accretor and outshines the secondary. This could have important implications for binary searches that expect Doppler modulation, as these typically rely on the fast-moving secondary to outshine the primary. If this role-reversal occurs, Doppler modulation may instead diminish as merger is approached.

The remainder of this paper is organized as follows. In \S~\ref{sec:Setup}, we discuss our hydrodynamics code, our initial setups and post-processing procedures. In \S~\ref{sec:Results} we present our main results on cavity morphologies, accretion rates, light curves, and periodicities, comparing systems with different mass ratios. Finally, in \S~\ref{sec:Con} we summarize our main conclusions, and discuss the consequences of our results, and  observational prospects.

\section{Hydrodynamical Setup, Post-processing, and Models}
\label{sec:Setup}

\subsection{Hydrodynamical Setup}
\label{subsec:HydroSetup}

In this section, we provide a brief overview of the technical details of {\tt Sailfish} \citep[for full details, see][]{Westernacher-Schneider2022}.

We solve the vertically integrated Newtonian fluid equations, keeping the lowest non-trivial order in powers of $z/r$ under the conditions of a thin disk ($h/r\ll 1$) and mirror symmetry about $z=0$.\
 These equations read
\begin{eqnarray}
    \partial_t \Sigma + \nabla_j \left( \Sigma v^j \right) &=& S_{\Sigma} \label{eq:mass} \\
    \partial_t \left( \Sigma v_i \right) + \nabla_j \left( \Sigma v^j v_i + \delta^j_i \mathcal{P} \right) &=& g_i + \nabla_j \tau^j_i + S_{p, i} \label{eq:mom} \\
    \partial_t E + \nabla_j \left[ \left( E+\mathcal{P} \right) v^j \right] &=& v^jg_j + \nabla_j \left( v^i \tau^j_i \right) \nonumber\\
    &-& \dot{Q} + S_{E} \label{eq:en},
\end{eqnarray}
where $\Sigma$ is the surface density, $\mathcal{P}$ is the vertically-integrated pressure, $v^i$ is the mid-plane horizontal fluid velocity, $E=\Sigma \epsilon + (1/2)\Sigma v^2$ is the vertically-integrated energy density, $\epsilon$ is the specific internal energy at the mid-plane of the disk, $g_i$ is the vertically-integrated gravitational force density, and $\tau^j_i = \Sigma \nu \left( \nabla_i v^j + \nabla^j v_i - (2/3)\delta^j_i \nabla_k v^k\right)$ is the viscous stress tensor (in a form that is trace-free in a 3-dimensional sense) with zero bulk viscosity, $\nu$ is the kinematic shear viscosity, and $S_{\Sigma}$, $S_{p,i}$, and $S_E$ are mass, momentum, and energy sinks, respectively. $\dot{Q}$ is the local blackbody cooling prescription, assuming hydrogen dominates the gas density \citep[see e.g][]{Frank2002}, given as

\begin{equation}
    \Dot{Q} = \frac{8}{3} \frac{\sigma} {\kappa\Sigma} \left(\frac{m_p\mathcal{P}}{k_B\Sigma}\right)^4 \label{eq:cooling},
\end{equation}

\noindent
where $\sigma$ is the Stefan-Boltzmann constant, $\kappa$ = 0.4 cm$^2$ g$^{-1}$ is the opacity due to electron scattering, $m_p$ is the proton mass, and $k_B$ is the Boltzmann constant. Thermal conductivity is neglected.

We initialize the disk to the conditions described in \cite{Goodman2003}:

\begin{align}
    \Sigma &= \Sigma_0\left(\frac{\sqrt{r^2+r_\mathrm{soft}^2}}{a_0}\right)^{-3/5}\\
    \mathcal{P} &= \mathcal{P}_0\left(\frac{\sqrt{r^2+r_\mathrm{soft}^2}}{a_0}\right)^{-3/2}\\
    \vec{v} &= \left({\frac{GM_{\rm bin}}{\sqrt{r^2+r_\mathrm{soft}^2}}}\right)^{1/2}\hat{\phi},
\end{align}

\noindent
where $G$ is the gravitational constant, $M_{\rm bin}$ is the total mass of the binary, $a_0$ is the initial binary separation, and $r_{\mathrm{soft}}$ is the gravitational softening length, which mimics the softening of the horizontal gravitational force that occurs due to vertical integration. This value is set equal to $r_{\rm s,bh}$ (where $r_{\rm s, bh} \equiv 2GM_{\rm bh}/c^2$ is the Schwarzschild radius of each BH, $M_{\rm bh}$ is the mass of an individual black hole, and $c$ is the speed of light). We also initialize a cavity at $r=2\,a_0$ by multiplying both $\Sigma$ and $\mathcal{P}$ by a window function $f(r)$, given by:
\begin{equation}
    f(r) = 10^{-4} + (1-10^{-4})\mathrm{exp}({-2a_0/(r^2+r_\mathrm{soft}^2})^{1/2})^{30}.
\end{equation}
We use a Shakura-Sunyaev viscosity prescription $\nu=\alpha c_{\rm s}h$ \citep{Shakura1973} with $\alpha=0.1$, and a $\Gamma$--law equation of state $\mathcal{P}=\Sigma\epsilon(\Gamma-1)$, where $\Gamma=5/3$. Both are consistent with the choices made in the fiducial model of \cite{Krauth2023}. The sound speed is given by $c_{\rm s}=\sqrt{\Gamma\mathcal{P}/\Sigma}$. The disk mass is small enough compared to the mass of the binary that its self-gravity can be ignored.

$\Sigma_0$ and $\mathcal{P}_0$ are chosen in such a way to adjust the characteristic disk aspect ratio $h/r$ to that which would be found in disks that include radiation pressure. But since radiation pressure is omitted in our thermodynamic models, if one interprets such disk aspect ratios literally, this implies extremely super-Eddington accretion rates~\citep{Dorazio2013}. This super-Eddington accretion rate should be viewed as an artificial aspect, whose only purpose is to yield characteristic aspect ratios consistent with those expected in the ``target'' radiation-dominated models. The correspondingly extreme luminosities are adjusted back down to the realistic range in post-processing (described more below). In the simulations, heating and cooling balance, and the disk aspect ratio develops self-consistently and locally. The effective temperature of the disk, $T_{\mathrm{eff}}$, is related to mid-plane temperature $T$ by
\begin{equation}
    T_{\mathrm{eff}}^{4} = \frac{4}{3}\frac{T^4}{\kappa\Sigma}
    \label{eq:teff}.
\end{equation}
and the accretion rate is approximately related to the  kinematic shear viscosity and density by $\dot{M}=3\pi\Sigma\nu$ \citep{Frank2002}.

The combined mass of the binary remains consistent with \cite{Krauth2023} at $M_{\rm bin}=10^6$~\(\rm M_\odot\), matching approximately the range at which LISA is most sensitive \citep{Amaro-Seoane2017}. We simulate three different mass ratios of $q = 0.1, 0.3$ and $0.5$. 

The binary separation during inspiral follows the quadrupole approximation of 
\cite{Peters1964}
\begin{equation}
    a(t) = a_0(1-t/\tau)^{1/4}
    \label{eq:peters},
\end{equation}
\noindent
where $\tau = a_0^4/4\beta$ is the total GW inspiral time and $\beta\equiv(64/5)(G^3/c^5)m_1m_2(m_1+m_2)$. The initial separation of the system is informed by the GW inspiral time and our desire to traverse the decoupling limit. While the viscous time (defined by $t_{\rm \nu} = 2/3\ r^2 / \nu$) of about 200 orbits at $r=a_0$ remains comparable in each simulation, the GW inspiral time increases with decreasing $q$. As such, this enables us to reduce the initial separation of the binary as we decrease $q$. Specifically, we chose an initial separation of 14, 18, and 20 $r_{\rm s,bin}$ (where $r_{\rm s, bin} \equiv 2GM_{\rm bin}/c^2$ is the Schwarzschild radius for the total mass of the binary) for $q = 0.1, 0.3$ and $0.5$, respectively. This will roughly cover the last $226$, $218$, and $250$ orbits of each system. Once the BH's event horizons make contact, the BHs are instantly merged, with a new position at their center of mass (the origin) with a new sink radius equal to the additive sink radii of the component masses. The BHs studied have zero spin and follow circular orbits.

Consistent with \cite{Krauth2023}, the gravitational field of each individual BH is modeled by a Plummer potential,
\begin{equation}
    \Phi_n = -\frac{GM_n}{\sqrt{r_n^2+r_{\mathrm{soft}}^2}},
\end{equation}
where $M_n$ is the mass of the $n$th BH, $r_n$ is the distance from a field point to the $n$th black hole, and $r_{\rm soft}$ is the gravitational softening length scale introduced above.

We use a torque-free sink prescription \citep{Dempsey2020, Dittmann2021} to model the removal of gas by each point mass. The sink radius is set to $r_{\rm s,bh}$ for each black hole. We explore the consequences of these choices in Appendix~\ref{app-a}. We use the secondary sink radius to set the spatial resolution for each simulation. We chose the resolution of each simulation to have the same number of cells across the sink radius of the smaller, secondary BH as for the equal-mass BHs in \cite{Krauth2023}. This ensures that we have consistent resolution near the BHs where it is needed most. We use a uniform Cartesian grid with a square domain of side length $16~a_0$. Because the radius of the secondary BH decreases with $q$, we need higher resolutions for lower mass ratios. Specifically, we use grid sizes of $4928^2$, $2496^2$, and $1920^2$ cells, and resolutions of $\Delta x=\Delta y\sim0.003~a_0$, $0.006~a_0$, and $0.008~a_0$, for $q = 0.1, 0.3$ and $0.5$, respectively.

As the realistic state of the disk, including the lopsided cavity and lump in the cavity wall, takes time to develop from our idealized axisymmetric initial conditions, we first run our simulations at lower resolution and without an inspiraling binary orbit. This allow a relaxed state of the gas to be established. We subsequently double the grid resolution multiple times. In practice, our first resolution in each case is about 1/3 of the final $\Delta x=\Delta y$ resolution, which we run for approximately 5 viscous times (650 initial orbits). We then double the resolution, and run for another full viscous time (130 initial orbits). Finally, we double the resolution and allow the fully evolved system to settle for an additional $\sim$third of a viscous time (40 initial orbits) before initializing inspiral.

\subsection{Post-processing}
\label{subsec:Post}

In post-processing, the effective temperature is obtained by combining Eqs.~\eqref{eq:cooling} and~\eqref{eq:teff}, leading to

\begin{equation}
    \Dot{Q}=2\sigma T_{\mathrm{eff}}^{4},
\end{equation}
\noindent
where the factor of 2 comes from the fact that the disk cools through two faces (top and bottom). The effective temperature relates to the accretion rate via
\begin{eqnarray}
    T_{\rm eff}^4 \propto \dot{M}.
\end{eqnarray}
The artificially high accretion rates discussed above also result in artificially high effective temperatures, which affect the EM luminosity and spectrum.  Therefore, we correct for this in post-processing by re-scaling the effective temperature back down to our target system via the map $T_{\rm eff}^4 \rightarrow T^4_{\rm eff} / M_{\rm boost}$, where $M_{\rm boost}$ is the dimensionless ratio of the accretion rate of the gas pressure model to that of the ``target'' model for which radiation pressure would be accounted for.

We assume blackbody emission from each cell of our domain, which enables us to compute the luminosity in different bands. We neglect Doppler effects, so our light curves are valid for observers who are viewing the disk sufficiently close to face-on. We calculate the luminosity of an area element $dA$ in the frequency band between $\nu_1$ and $\nu_2$ via
\begin{equation}
    dL=\pi dA \int_{\nu_1}^{\nu_2}\frac{2h\nu^3/c^2}{ \mathrm{exp}\left(\frac{h\nu}{kT_\mathrm{eff}}\right)-1 }d\nu.
\end{equation}
To obtain the total luminosity of the disk, we sum over the spatial domain. Both \cite{Krauth2023} and \cite{Franchini2024} showed that the luminosity in the optical band originates from the CBD and remains approximately unchanged through merger. In this work, we saw the same behavior, and as such we chose to focus on luminosities in the two fixed bands of $E_{\rm UV}:3.1-124.0~\rm eV$ and $E_{\rm X-ray}:0.124-124.0~\rm keV$.

\section{Results}
\label{sec:Results}

In this section, we describe the results from our suite of simulations and compare them to the fiducial equal-mass model of \cite{Krauth2023}. We begin by comparing the disk morphologies between the models. We then present the accretion rates, followed by light curves in order of decreasing mass ratio, to discuss what changes occur as the binary mass ratio decreases. We discuss what implications our findings may have on searches involving Doppler modulations. Finally, we examine the pre-merger periodicities of these systems. The relevant figures from \cite{Krauth2023} have been included for convenience.

\subsection{Cavity morphologies}
\label{subsec:morphs}

\begin{figure*}
    \centering
    \includegraphics[width=0.95\textwidth]{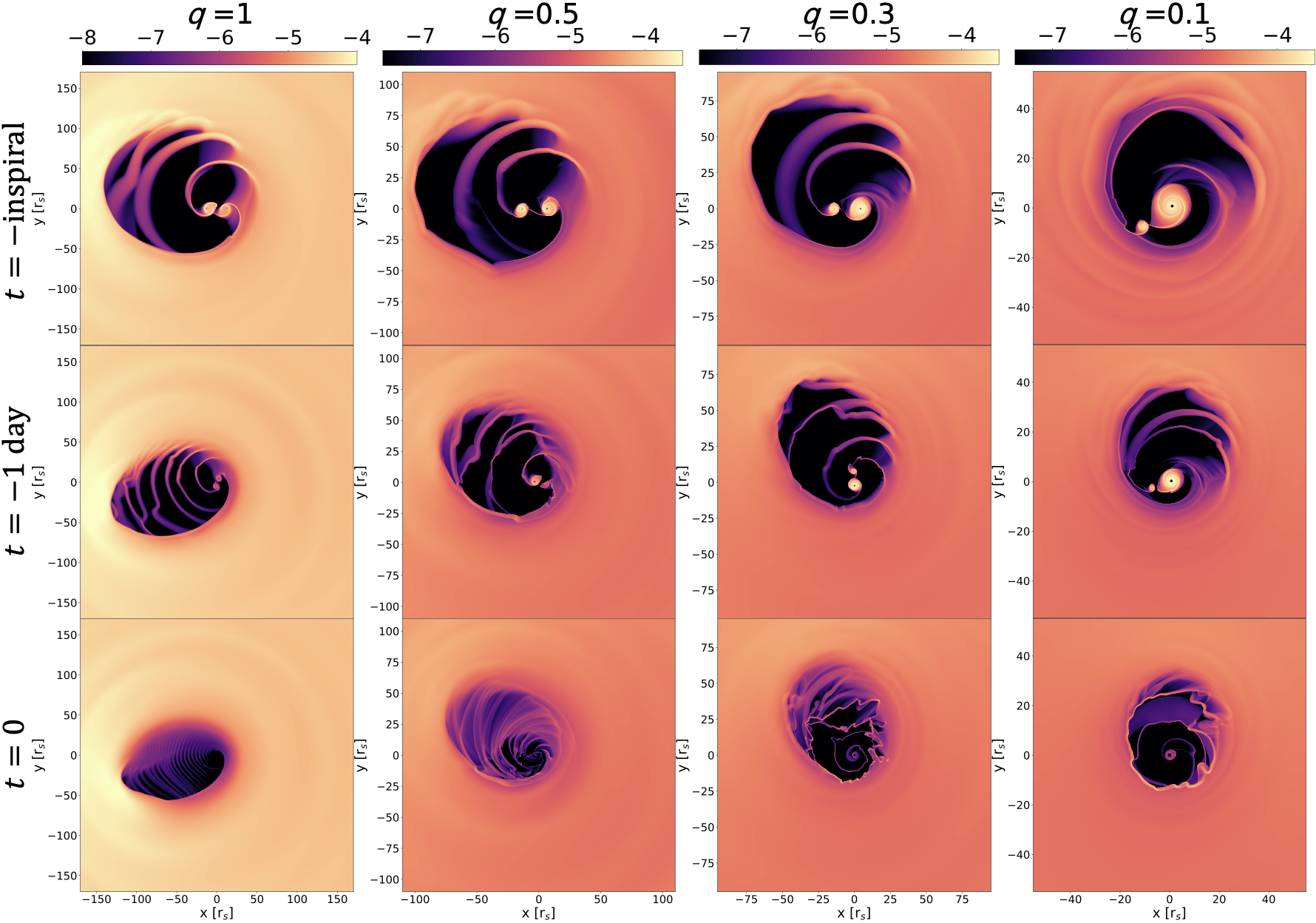}
    \caption{Zoomed-in snapshots of the logarithmic surface density. From left to right, we show models $q=1$ \protect\citep[from][]{Krauth2023}, $q=0.5$, $q=0.3$, and $q=0.1$ (from this work). The total mass of the binary $M_1+M_2$ is the same in all cases. From top to bottom, we show snapshots at different times: at the initiation of inspiral, 1 day before merger, and at merger.}
    \label{fig:comb_dens}
\end{figure*}

Figure \ref{fig:comb_dens} shows zoomed-in snapshots of the logarithmic surface density for all mass ratio models. The first column is adapted from \cite{Krauth2023} for their $q=1$ runs. The second, third, and fourth columns correspond to our $q=0.5$, $q=0.3$, and $q=0.1$ runs, respectively. Moving from top to bottom in each column, we show snapshots at the beginning of inspiral, 1 day before merger, and at merger. 

It is important to first point out that because these simulations were initiated at different binary separations, the plots shown here vary in scale in order to facilitate visual inspection (see the $x$- and $y$-axes). Thus, we caution the reader about visually comparing sizes across models in this figure. Nonetheless, several distinguishing trends are evident with decreasing $q$. First, we see the overall eccentricity of the cavity decreases with lower $q$, the most obvious change being visible in the $q=0.1$ case. Although the cavity is offset from the center of mass of the system, it is more circular than the other models. The cavity also decreases in size with smaller $q$. To somewhat remove the dependence on the initial separation, if we state the semimajor axis of the cavity in units of the initial semi-major axis of the binary, we find it decreases with $q$ as $3.6~a_{0, q=1}$, $3.2~a_{0, q=0.5}$, $3.0~a_{0, q=0.3}$, $2.9~a_{0, q=0.1}$.

Next, we also notice a fairly significant change in the ratio of the primary minidisk radius to that of the secondary with decreasing $q$. For circular binaries, this can be summarized by the fitting formulas for tidal truncation radii for unequal-mass binaries in the range of $0.01<q<1$ \citep{Eggleton1983,Roedig2014}:
\begin{align}
    R_{\rm prim} &\simeq 0.27q^{-0.3}a,
    \label{eq:md_rad_p} \\
    R_{\rm sec} &\simeq 0.27q^{0.3}a.
    \label{eq:md_rad_s}
\end{align}
Not only will the secondary minidisk have a much smaller radius, but as we decrease $q$, this ratio will grow according to the power-laws described by these equations. As we progress in time, on an absolute basis, the decrease in radius of the primary minidisk is hastened for lower $q$, whereas the decrease in the radius of the secondary's will be slowed down.

The amount of material in accretion streams flung back into the cavity wall decreases with $q$, likely because the primary at some point becomes massive enough that is effectively stationary. Although it can still partially create streams, the streams are slow-moving and have low density. The secondary becomes the main component to tidally strip material from the close edge of the cavity wall. It flings material with enough velocity to overtake the more minor streams created by the primary. This effectively halves the frequency of stream creation. As we progress in time, to the row 1 day before merger, this effect begins to also present at higher $q$. As the binary contraction hastens, the distance between the binary and the cavity wall increases. This further reduces the primary's ability to create streams.

For lower-$q$ systems, the burgeoning new single minidisk around the merger remnant is more prominent. It seems, as $q$ is decreased, the system is less proficient at tidally truncating the minidisks approaching merger, and there remains traces of the original minidisks (most notably the primary). That having been said, the merger remnant is increasingly far away from a new source of replenishment from the nearest side of the cavity wall, stranded without new accretion material nearby.

\subsection{Accretion rates}
\label{subsec:acc}

\begin{figure*}
    \centering
    \includegraphics[width=0.93\textwidth]{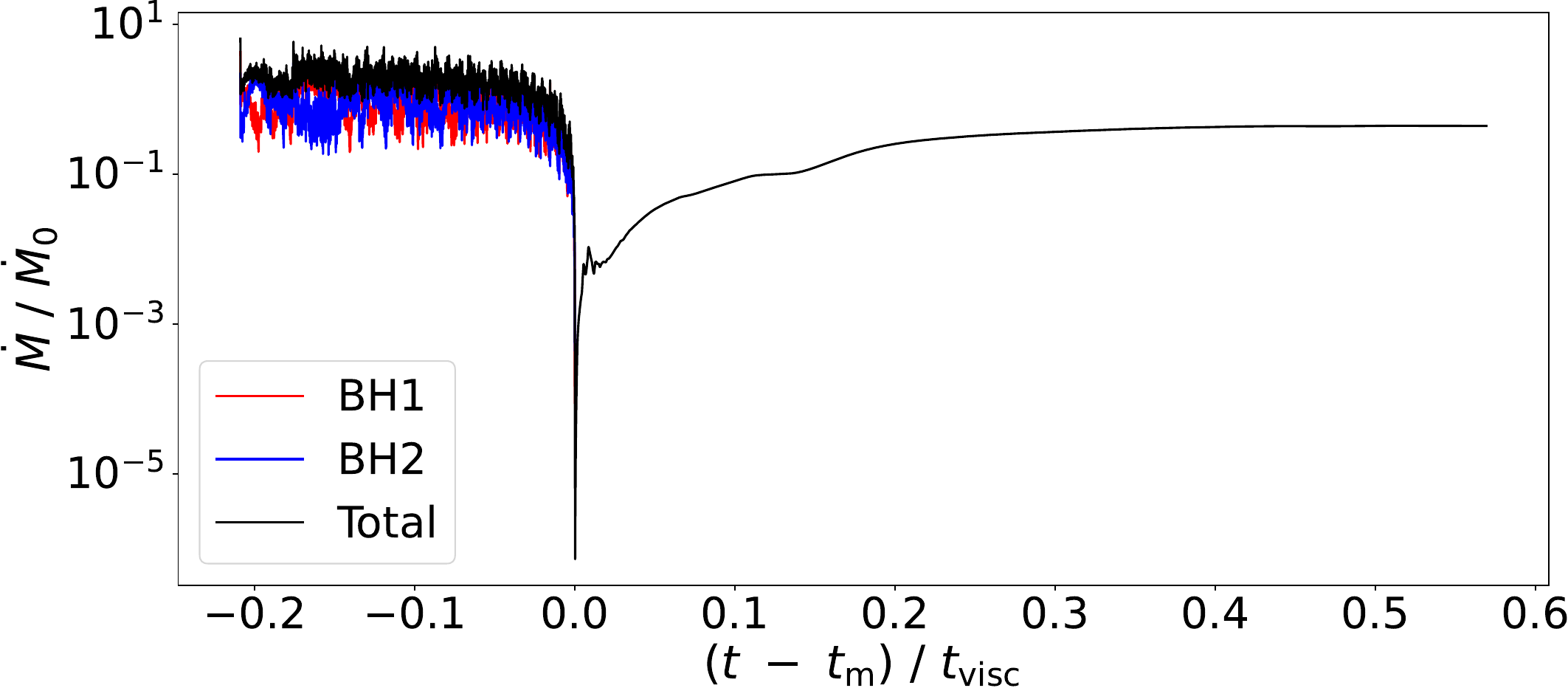}
    \caption{Black hole accretion rates in the $q=1$ fiducial model before and after the time of merger, $t_m$, for the case with no post-merger recoil or mass loss. Accretion rates are normalized by the steady Shakura-Sunyaev value around a single BH  \protect\cite[adapted from Fig. 2 in][]{Krauth2023}.}
    \label{fig:fid_long_term_accretion}
\end{figure*}

Here we compare the accretion rates from \cite{Krauth2023}'s $q=1$ fiducial model, and our $q=0.5$, $0.3$ and $0.1$ models. Starting with the $q=1$ case in Figure~\ref{fig:fid_long_term_accretion}, we see that neither BH dominates the accretion for all time, but they take turns doing so. The most striking feature is the large drop found approaching merger. This several order of magnitude drop is due to the tidal truncation of the minidisks as inspiral progresses, effectively dismantling the minidisks by the time of merger.

\begin{figure*}
    \centering
    \includegraphics[width=0.9\textwidth]{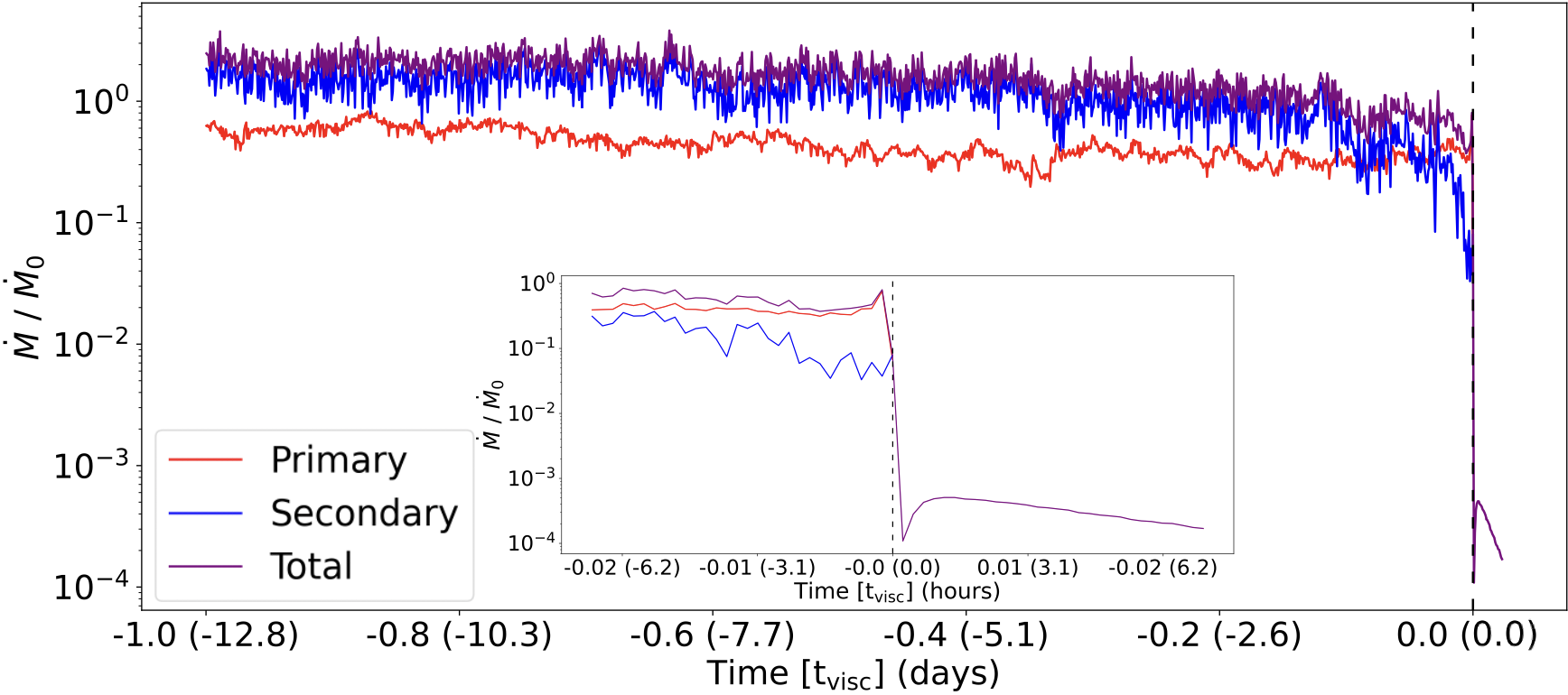}
    \caption{Model $q=0.5$: Black hole accretion rates $\sim1$ viscous time before the time of merger, $t_m$, and a short while after. The inset is zoomed-in on the time of merger. Accretion rates are normalized by the steady Shakura-Sunyaev value around a single BH. While we see the secondary initially out-accretes the primary, as we approach merger that dominance switches as the secondary force-feeds the primary. We see the accretion for both BHs begins dropping before merger occurs, and plummets several orders of magnitude into post-merger.}
    \label{fig:mod6_acc}
\end{figure*}

It is interesting to examine the pre-merger behavior for the $q<1$ cases, since they contrast with the $q=1$ case in a few ways. For the $q=0.5$ accretion rates, seen in Figure~\ref{fig:mod6_acc}, we notice the secondary now clearly out-accretes the primary, at least until the final moments. We can see in the last $\sim$tenth of a viscous time, the primary begins to accrete more than the secondary. By looking at the zoomed-in inset plot of this figure, we can actually see that the primary increases its accretion rate very shortly before merger. Nonetheless, in the final moments, we see both the secondary and the primary accretion rates plummet. Again, the post-merger accretion rates are several orders of magnitude lower than prior to merger. Note, however, that the drop is less dramatic than the $q=1$ case, indicating that while the minidisks are still dismantled, the process is less efficient with lower $q$. We believe that these findings are explained by the following two reasons. First, with decreasing $q$, the tidal influence of the secondary, and its ability to affect gas near the primary, is diminished. Second, in the later stages of inspiral, there appears to be a more preferential transfer of mass from the secondary to the primary. The secondary effectively ``force-feeds'' the primary. That is, after scooping material from the cavity wall, it appears to deliver a sizable fraction of this gas to the primary, thereby increasing the accretion rate of the primary.

\begin{figure*}
    \centering
    \includegraphics[width=0.9\textwidth]{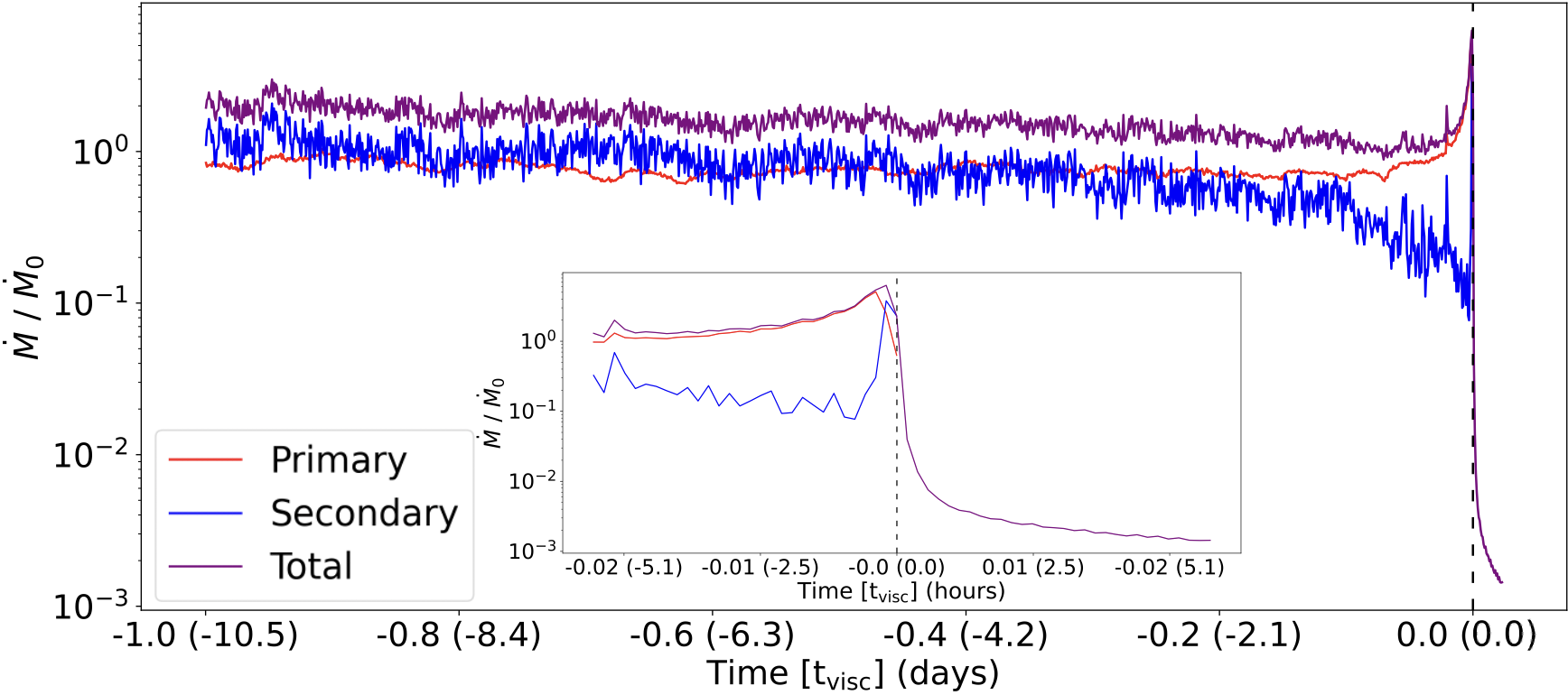}
    \caption{As in Fig.~\ref{fig:mod6_acc}, but for $q=0.3$: while we see the secondary initially out-accretes the primary, as we approach merger that dominance switches as the secondary force-feeds the primary. With decreased $q$, we see this feeding is even stronger with a sharper rise in the primary accretion rate pre-merger. Additionally, even the secondary has a brief but sharp spike before merger, showing increased accretion shortly before merger. However, we still see the accretion for both BHs begins dropping before merger occurs. It still plummets several orders of magnitude into post-merger, but less so than in the $q=0.5$ case.}
    \label{fig:mod5_acc}
\end{figure*}

Moving to Figure~\ref{fig:mod5_acc}, we see that progressing lower to $q=0.3$ only intensifies the features we saw for $q=0.5$. Again, earlier on during inspiral, the secondary accretes more than the primary. However, now the switch is more dramatic. Somewhere near $\sim 0.4$ viscous times remaining, their roles switch and the primary becomes and remains the dominant accretor. We also now see that the primary steeply increases its accretion by a factor of a few shortly before merger. Additionally, there is an order of magnitude spike in the accretion of the secondary moments just before merger as well. With even lower $q$, it seems both the lessened destruction, and the replenishment of the secondary minidisk from the cavity wall leading to a more pronounced force-feeding, not only sustains the accretion of the primary, but substantially increases it shortly before merger. Regardless, once again, the final moments of merger disperse the minidisks and cause both the primary and secondary accretion rates to begin to plummet just before merger, as shown in the zoomed-in inset. Additionally, post-merger, the recovery of the accretion rate after its several-orders-of-magnitude drop is not prompt. Note that the post-merger accretion rate in this $q=0.3$ case is slightly higher than in the $q=0.5$ case. 

\begin{figure*}
    \centering
    \includegraphics[width=0.9\textwidth]{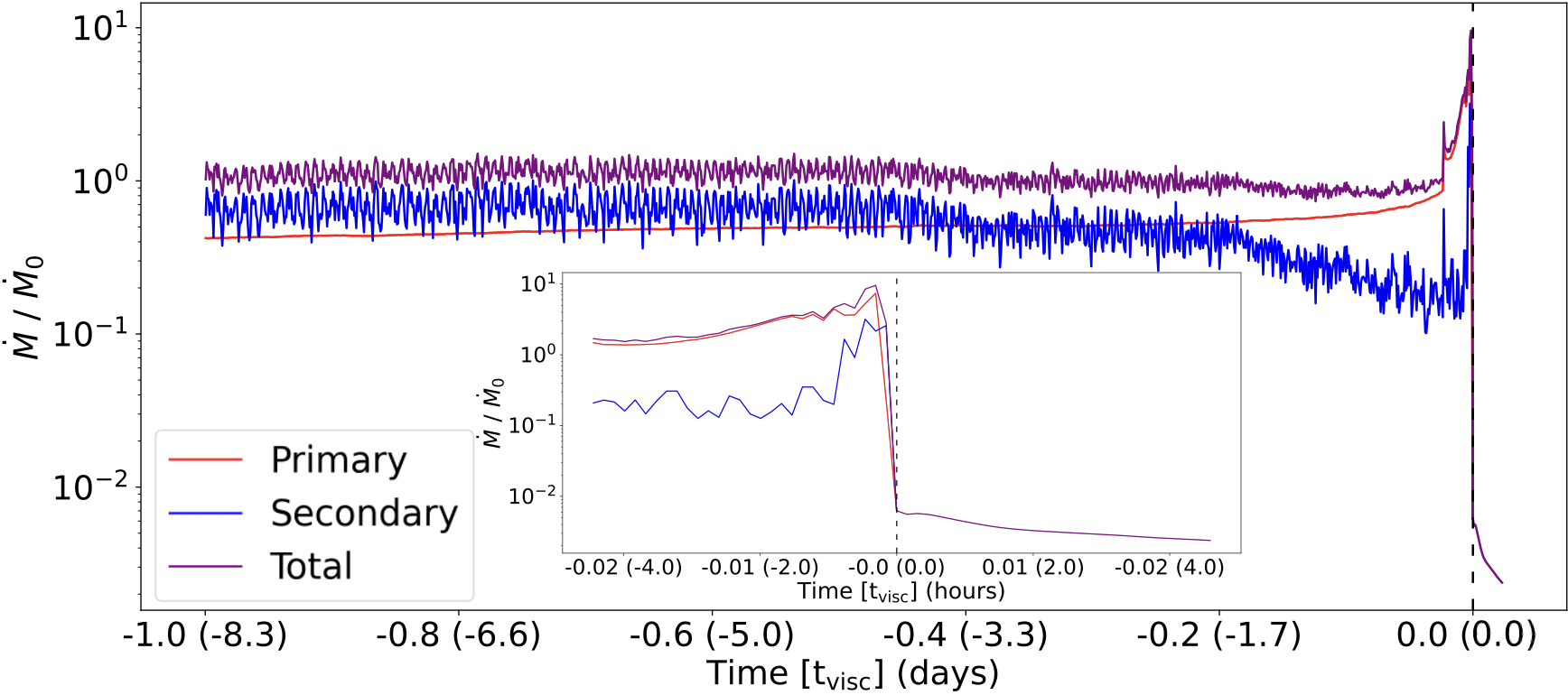}
    \caption{As in Fig.~\ref{fig:mod6_acc}, but for $q=0.1$: while we see the secondary initially out-accretes the primary, as we approach merger that dominance switches as the secondary force-feeds the primary. With further decreased $q$, we see this force-feeding is even more dramatic, once again with the secondary spiking shortly before merger. Still though, we see the accretion for both BHs begins dropping before merger occurs. Accretion again plummets several orders of magnitude into post-merger, but even less so than even the $q=0.3$ case now.}
    \label{fig:mod4_acc}
\end{figure*}

Figure~\ref{fig:mod4_acc} shows the accretion rates for our final case of $q=0.1$. We again see comparable but exacerbated features from the prior case. Once again we see a switch of the dominant accretor from the secondary to the primary near $\sim0.4$ viscous times remaining. It appears that further lessened dismantling and increased force-feeding from the secondary leads to an accretion rate that is even greater than before. The primary jumps even more than the $q=0.3$ case. The secondary increases a similar amount to the prior case, but does so even earlier on before merger. It should be noted that the final data point for the secondary accretion was so low ($\sim10^{-8}$) that it is off the $y$-axis range in Figure~\ref{fig:mod4_acc}. Accretion rates for both BHs still drop in the final moments before merger. Again, the post-merger accretion rate remains several orders of magnitude below the pre-merger total, although now the rate is slightly higher than the post-merger rate in the $q=0.3$ case. It seems with decreased $q$ there is a decreased drop in total accretion from pre- to post-merger, but that it regardless remains a several order of magnitude drop for the $q$ values we considered.

The role-reversal here may help reconcile some previous findings. For a system with unequal mass ratio, many studies have shown that in some binary systems, the secondary can accrete more efficiently or have more energetic emission properties than the primary \citep{Farris2014,Dorazio2015}. However, other papers have reported the primary as the dominant accretor at late inspiral \citep{Gold2014}. The findings here appear to bridge the gap between these previous results, and show there is a transition in the dominant accretor from the secondary to the primary as inspiral progresses. Literature investigating primary- vs secondary-dominated accretion in binaries suggests that which accretor dominates is dependent upon cooling rates, or simply how hot the minidisks are, and perhaps the strength of viscosity \citep[see e.g.][and references therein]{Hanawa+2010,Ochi+2005,Young+2015,Bourne+2024}. Whether these mechanisms explain the transition we find during inspiral is left to future work.

\subsection{Light curves}
\label{subsec:lcs}

We now compare the light curves for the equal-mass fiducial model of \cite{Krauth2023} and our unequal-mass simulations in order of decreasing $q$.

\begin{figure*}
    \centering
    \includegraphics[width=0.9\textwidth]{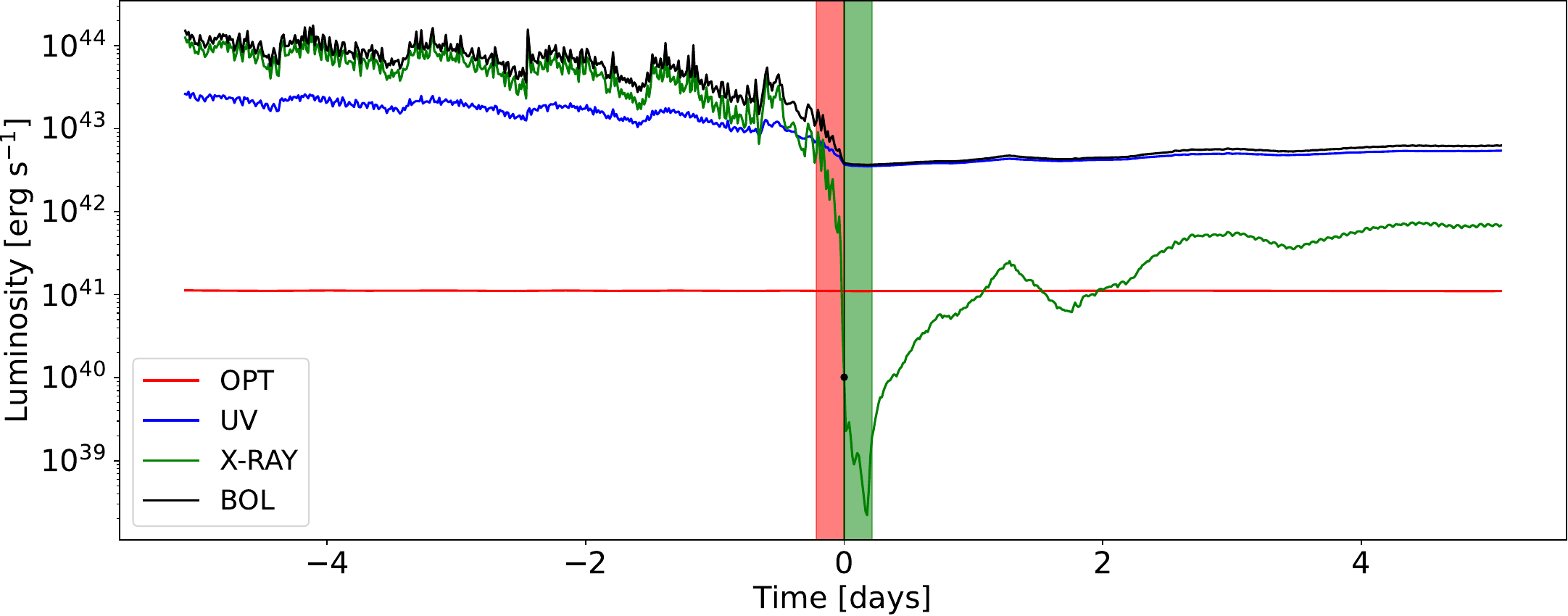}
    \caption{Optical, UV, X-ray, and bolometric light curves from five days before merger to five days after merger $q=1$ fiducial model. The black vertical line is centered at the merger ($t=0$), with a black dot marking the X-ray luminosity at the moment of the merger. The red/green zones indicate five hours before/after the merger. The inset shows a zoom-in version around the merger.  This run includes a recoil kick contributing to the post-merger recovery of the luminosity, but does not include mass loss, which inhibits recovery \protect\cite[adapted from Fig. 4 in][]{Krauth2023}.}
    \label{fig:fid_bands_5days}
\end{figure*}

Beginning with the equal-mass case in Figure~\ref{fig:fid_bands_5days}, we notice defined periodicity in the UV and X-ray light curves on the timescale of the orbital period for the lump days before merger. Most notable is the several order of magnitude drop in the X-ray luminosity starting just a handful of hours before merger. Post-merger includes a recoil of the binary remnant which boosts the luminosity, but regardless the system stays X-ray dark for several days post-merger. The mechanism responsible for this is the aforementioned tidal truncation of the minidisks by the binary potential. Since the minidisks dominate the X-ray emission and they survive up until hours before the merger, their destruction leads to a corresponding sharp drop in X-ray emission.

\begin{figure*}
    \centering
    \includegraphics[width=0.9\textwidth]{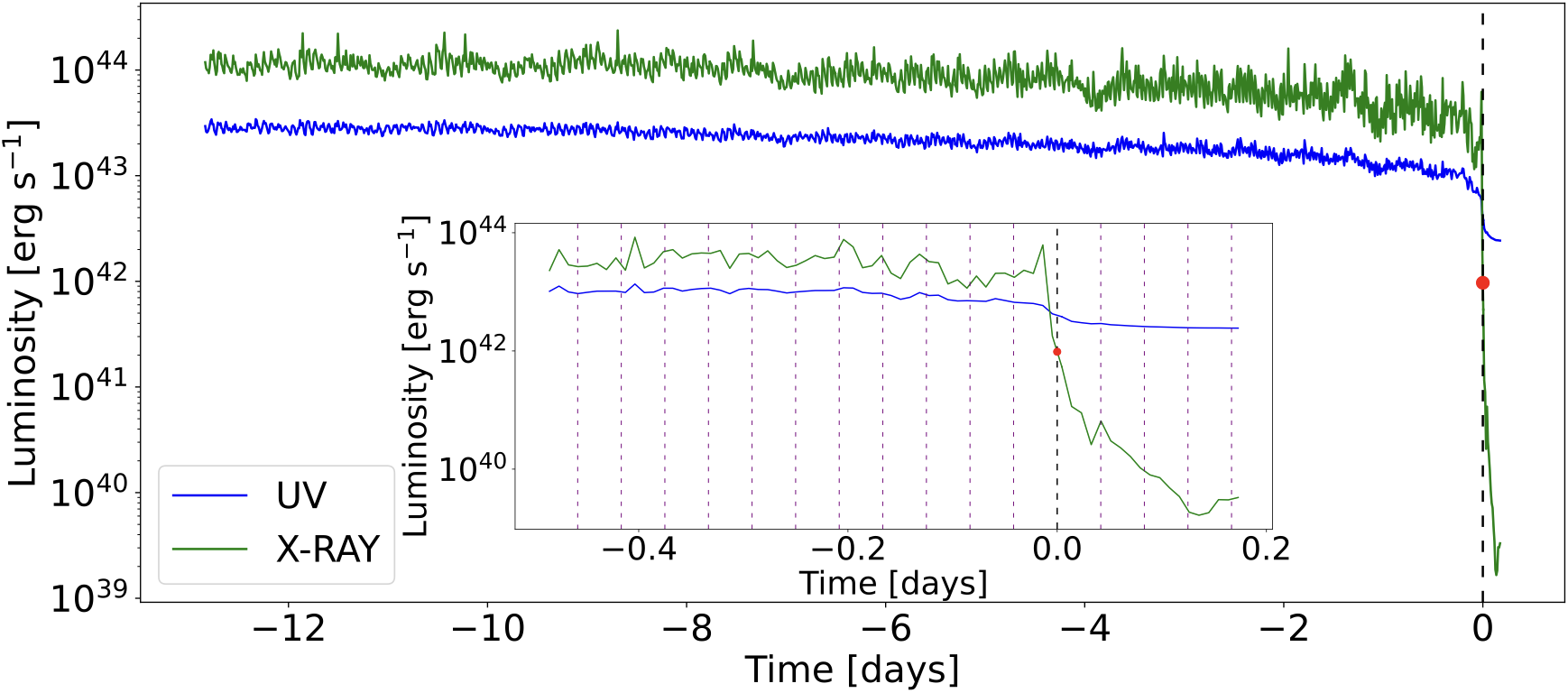}
    \caption{Model $q=0.5$: UV (blue) and X-ray (green) light curves for the inspiral  before merger to several hours after merger with a zoomed-in inset for the final $\sim$half a day until after merger. The black dashed vertical line marks the moment of merger, with a red dot marking the X-ray luminosity at the moment of the merger. The purple dashed lines in the inset mark 1 hr intervals. We see a slight decline in both luminosity bands premerger, with a sharp drop in the X-ray luminosity (and a smaller drop in UV luminosity) from pre- to post merger. This can be seen most clearly in the inset, which shows this drop begins 10s of minutes before merger.}
    \label{fig:mod6_lc}
\end{figure*}

Looking at Figure~\ref{fig:mod6_lc}, we can see the effect of decreased $q$. In addition to the larger window displaying the UV (blue) and X-ray (green) luminosities for the full inspiral, the smaller inset window zooms in on the last $\sim$half day of inspiral and post-merger. We see that decreased $q$ values result in several notable modifications to the light curves. There is a slight overall decline in the luminosity during inspiral. Once again, the tell-tale feature found in \cite{Krauth2023} of the several order of magnitude drop in thermal X-ray luminosity from pre- to post-merger is seen, but it has been modified; the drop now occurs much later, in the final 10s of minutes, instead of the hours before, as seen in the equal-mass case. Immediately after merger, the X-ray luminosity continues to decline, reaching comparable values to that of the equal-mass case. Again, the system stays X-ray dark for hours after merger. The UV luminosity also decreased slightly, but again less so than in the equal-mass case.

\begin{figure*}
    \centering
    \includegraphics[width=0.9\textwidth]{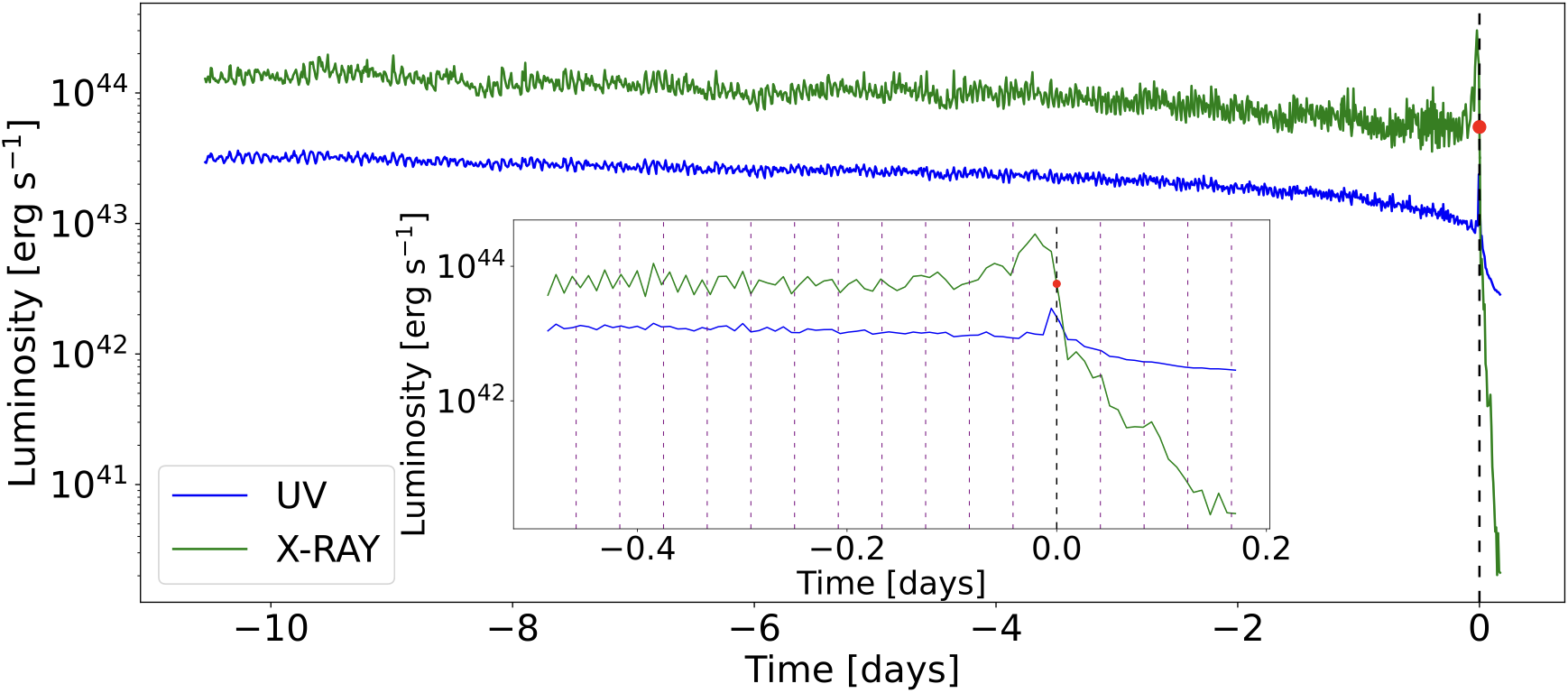}
    \caption{As in Fig.~\ref{fig:mod6_lc}, but for $q=0.3$: there is again a slight decline in both luminosity bands pre-merger, with a sharp drop in the X-ray luminosity (and a smaller drop in UV luminosity) from pre- to post merger. The drop again begins 10s of minutes before merger, but now there is also a sharp rise in the luminosity just before before the drop occurs.}
    \label{fig:mod5_lc}
\end{figure*}

As we decrease $q$ further, we actually see an entirely new, distinguishing feature emerge. We show the light curves for the $q=0.3$ model in Figure~\ref{fig:mod5_lc}. There is again a small overall decline in the luminosity throughout inspiral until near merger. Additionally, before the X-ray drop, there is a steep increase, or flare. While the system once again undergoes a drop by several orders of magnitude in the thermal X-ray luminosity as it traverses merger, there is an additional factor-of-several increase in the thermal X-ray luminosity before this occurs. This increase appears to begin on the timescale of hours before merger. We see the X-ray luminosity then drops to a comparable value to that before the spike, and continues falling dramatically. This bump and dip can also be seen to a lesser degree in the UV light curves as well. These jumps occur as the increase in accretion rates, seen in Figure~\ref{fig:mod5_acc} becomes most significant. The decreased destruction and increased force feeding of the primary from the secondary appears to not only increase the accretion rates of (most notably) the primary shortly before merger, but also creates a more excited system, leading to increased thermal X-ray luminosity before the drop occurs.

\begin{figure*}
    \centering
    \includegraphics[width=0.9\textwidth]{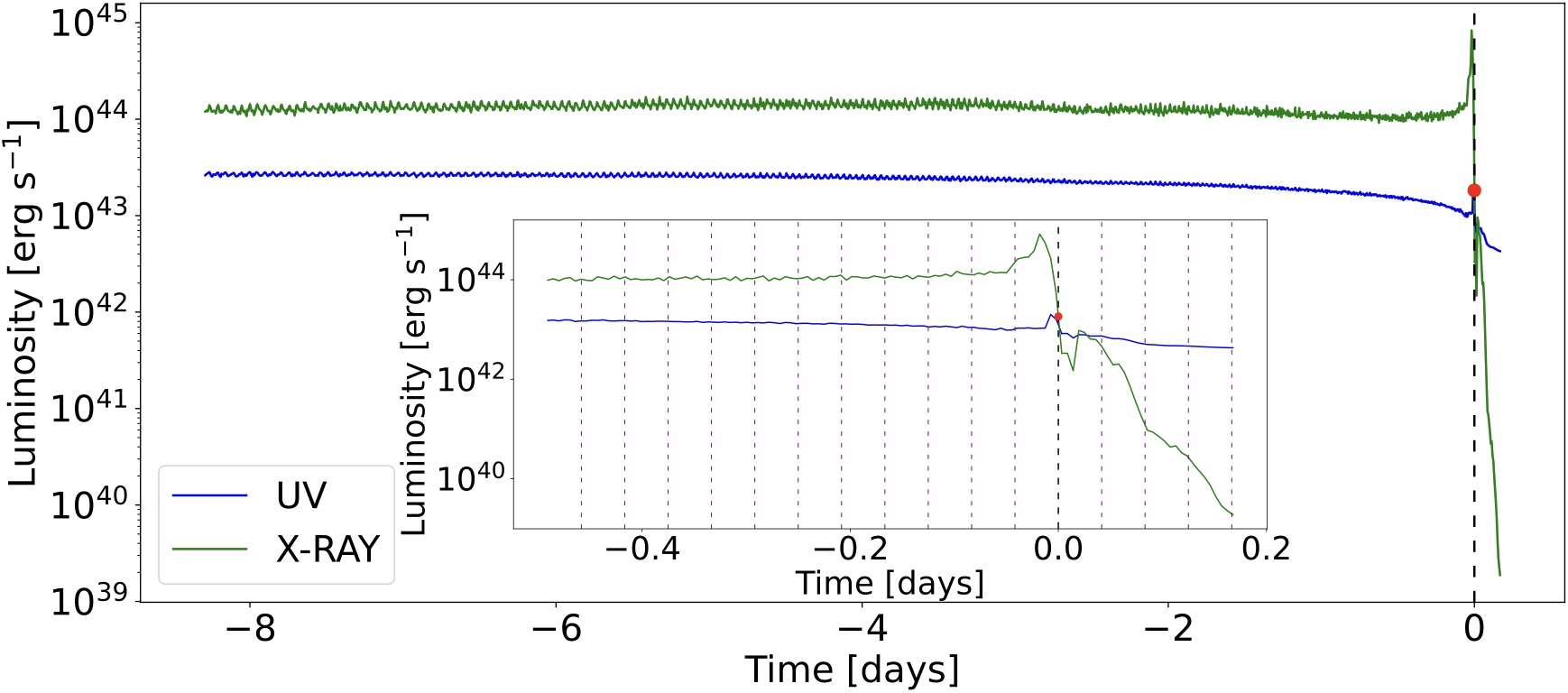}
    \caption{As in Fig.~\ref{fig:mod6_lc}, but for $q=0.1$: the decline in the X-ray luminosity pre-merger have essentially plateaued, with only a small overall decline to the UV in late inspiral. We again see a several order of magnitude drop in the X-ray luminosity (and a smaller drop in UV luminosity) from pre- to post-merger with a sharp rise just before this drop. The rise is now even even more pronounced and begins sooner.}
    \label{fig:mod4_lc}
\end{figure*}

Finally, we examine the light curves of the lowest mass ratio case of $q=0.1$ in Figure~\ref{fig:mod4_lc}. The overall decline in luminosity during inspiral has mostly vanished, now staying at nearly the same level throughout inspiral until merger. Once again the system undergoes a several order of magnitude drop in the thermal X-ray luminosity from pre- to post-merger. Now, however, the pre-merger silence is interrupted with a larger, near order of magnitude spike in X-ray luminosity before the drop occurs, once again on the order of hours before merger. Post-merger, the system again remains X-ray dark. This spike and dip is once again seen to a lesser degree in the UV band. These spikes again correspond to the increased accretion rates of the $q=0.1$ model seen in Figure~\ref{fig:mod4_acc}. Now we see the further lessened destruction and enhanced forceful feeding from the secondary leads to increased X-ray brightness of the system before the tell-tale drop in X-ray luminosity $\sim$hours before merger. A pre-merger flare like this was previously suggested by \cite{Chang2010}, but attributed to a different mechanism known as ``tidal squeezing'' of the primary minidisk by the secondary. However, this explanation is disfavored by dynamical considerations and simulations, which indicate that the tidal torque from the secondary before the merger is too weak to reduce the orbits of primary disk gas parcels faster than the binary's contraction \citep{Clyburn2024}. As this pre-merger flare could potentially be used as a new identifying signature, we test its robustness with sensitivity tests in Appendix~\ref{app-a}.

\subsection{Implications for Doppler modulation}
\label{subsec:dopmods}

For a system with unequal mass ratio, studies have shown that in most binary systems, the secondary accretes more efficiently or has more energetic emission properties \citep{Farris2014,Dorazio2015,Shi2015,Munoz2020,Duffell2020,Siwek2020,Siwek2023}. Additionally, because the secondary's velocity scales with $1/(1+q)$, the secondary moves faster for unequal masses, so it has a larger Doppler amplitude \citep{Dorazio2015}. As such, for standard spectral slopes of $\alpha_{\nu}\equiv d\ln F_\nu /d\ln\nu<3$, previous studies have proposed that the binary will appear brighter (dimmer), when the secondary is approaching (receding) from the observer, even if the rest-frame luminosity is constant.

Given the role-reversal in accretion rates of the secondary and primary approaching merger that we find with decreased $q$, in this section we also examine the individual contributions of each minidisk to the total X-ray light curve during inspiral. We do so by calculating the emission in a circular zone centered at each BH individually, with their radii scaling in accordance with Eqs.~\ref{eq:md_rad_p} and \ref{eq:md_rad_s}.

\begin{figure}
    \centering
    \includegraphics[width=0.45\textwidth]{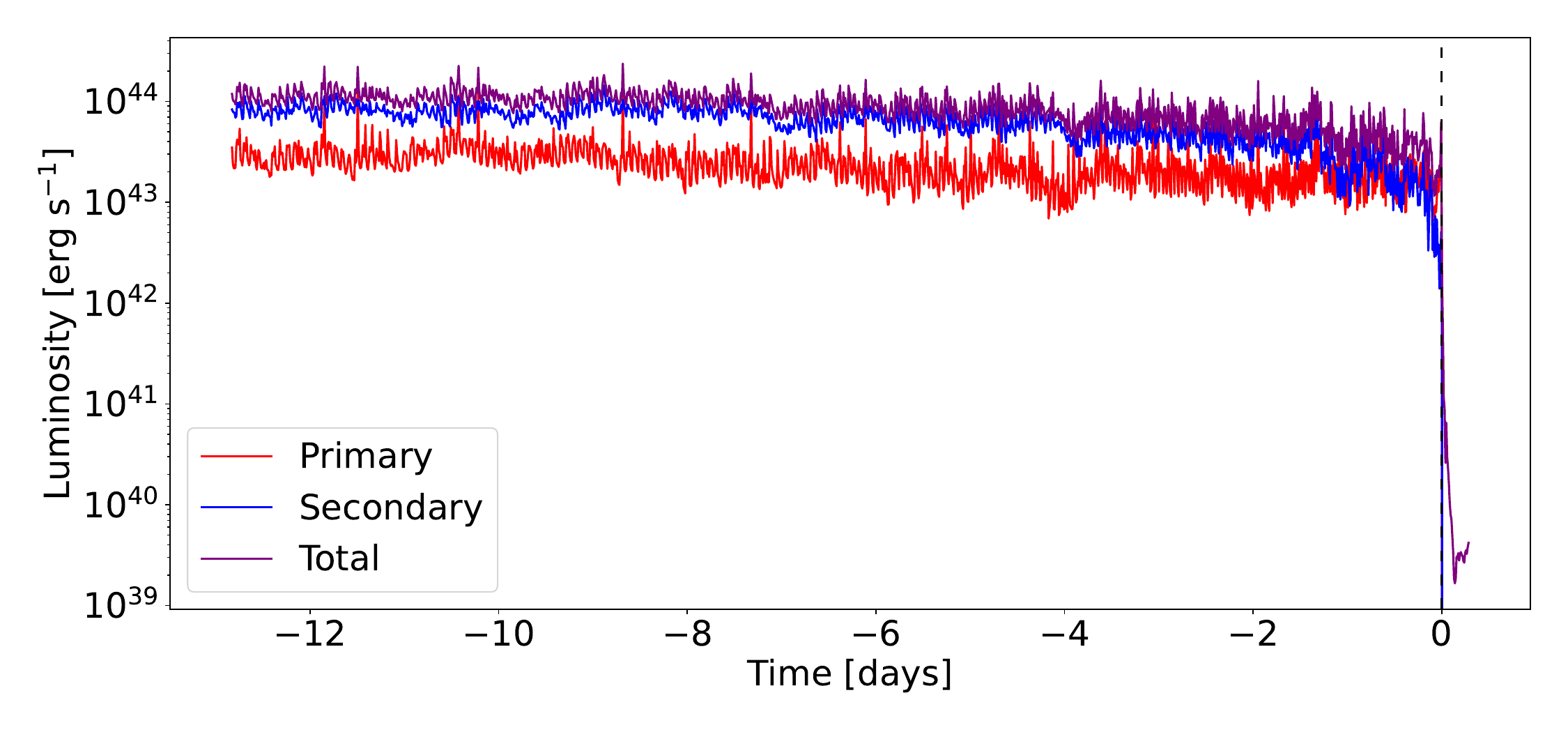}
    \caption{Model $q=0.5$: X-ray light curves for the inspiral to through merger. We see the primary (red) contribution to the overall X-ray luminosity is less than the secondary (blue) contribution until the final moments before merger.}
    \label{fig:mod6_mds}
\end{figure}

First, Figure~\ref{fig:mod6_mds} shows the total X-ray light curve and each minidisks' contribution for the $q=0.5$ case. We see that the secondary (blue) dominates the X-ray luminosity leading to merger, until in the final moments, the primary outshines the secondary at around the same time the role-reversal in accretion rates occurs.

\begin{figure}
    \centering
    \includegraphics[width=0.45\textwidth]{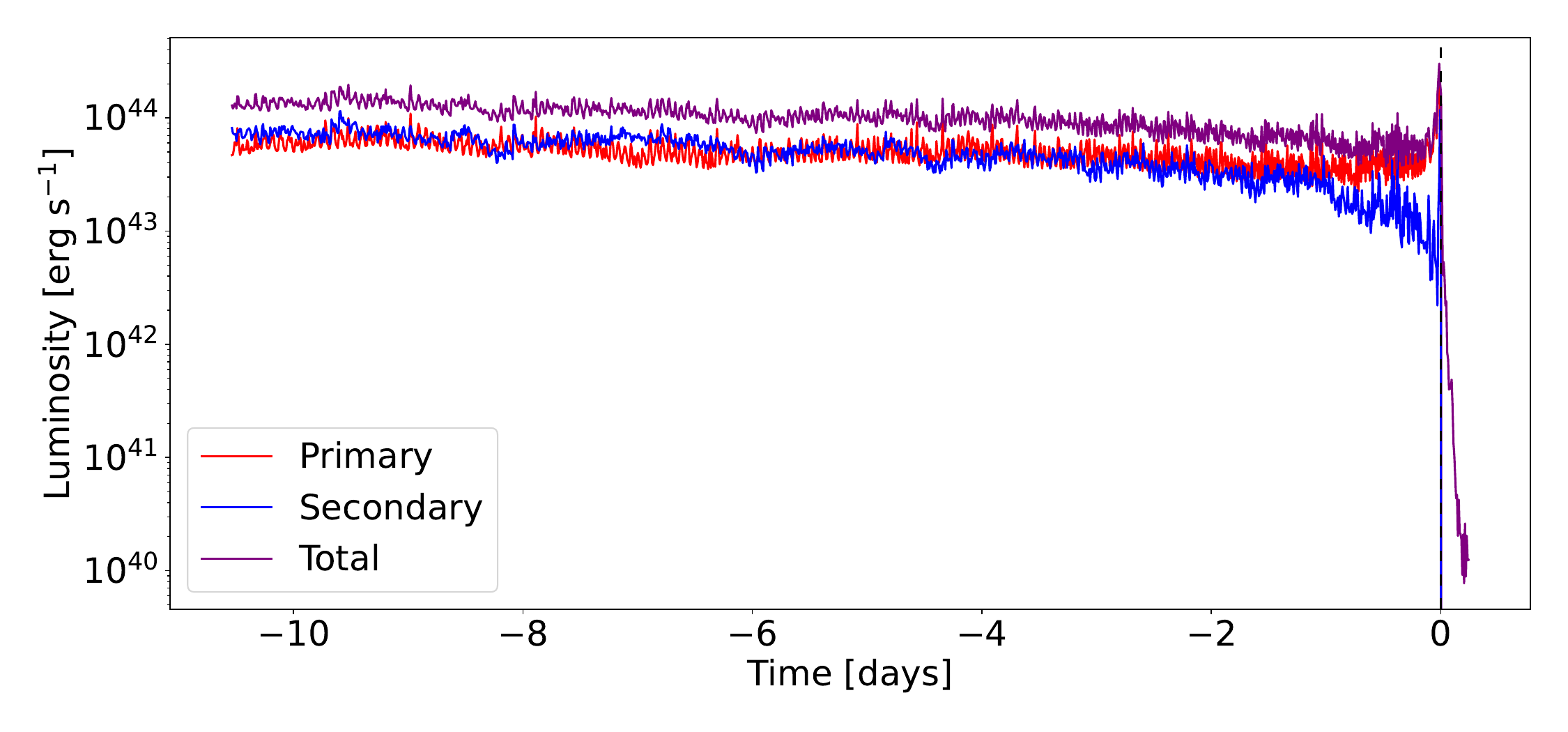}
    \caption{Model $q=0.3$: X-ray light curves for the inspiral to through merger. Early inspiral, we see nearly equal contribution to the overall luminosity from the primary (red) and the secondary (blue). However, in the last days of inspiral, the primary begins to outshine the secondary.}
    \label{fig:mod5_mds}
\end{figure}

Next, Figure~\ref{fig:mod5_mds} shows the $q=0.3$ case. We see that early on during inspiral, the primary and secondary are both contributing to the total X-ray luminosity by comparable amounts. However, as inspiral progresses, we see once again that when the primary becomes the dominate accretor, the primary begins to dominate the X-ray luminosity in the final days before merger.

\begin{figure}
    \centering
    \includegraphics[width=0.45\textwidth]{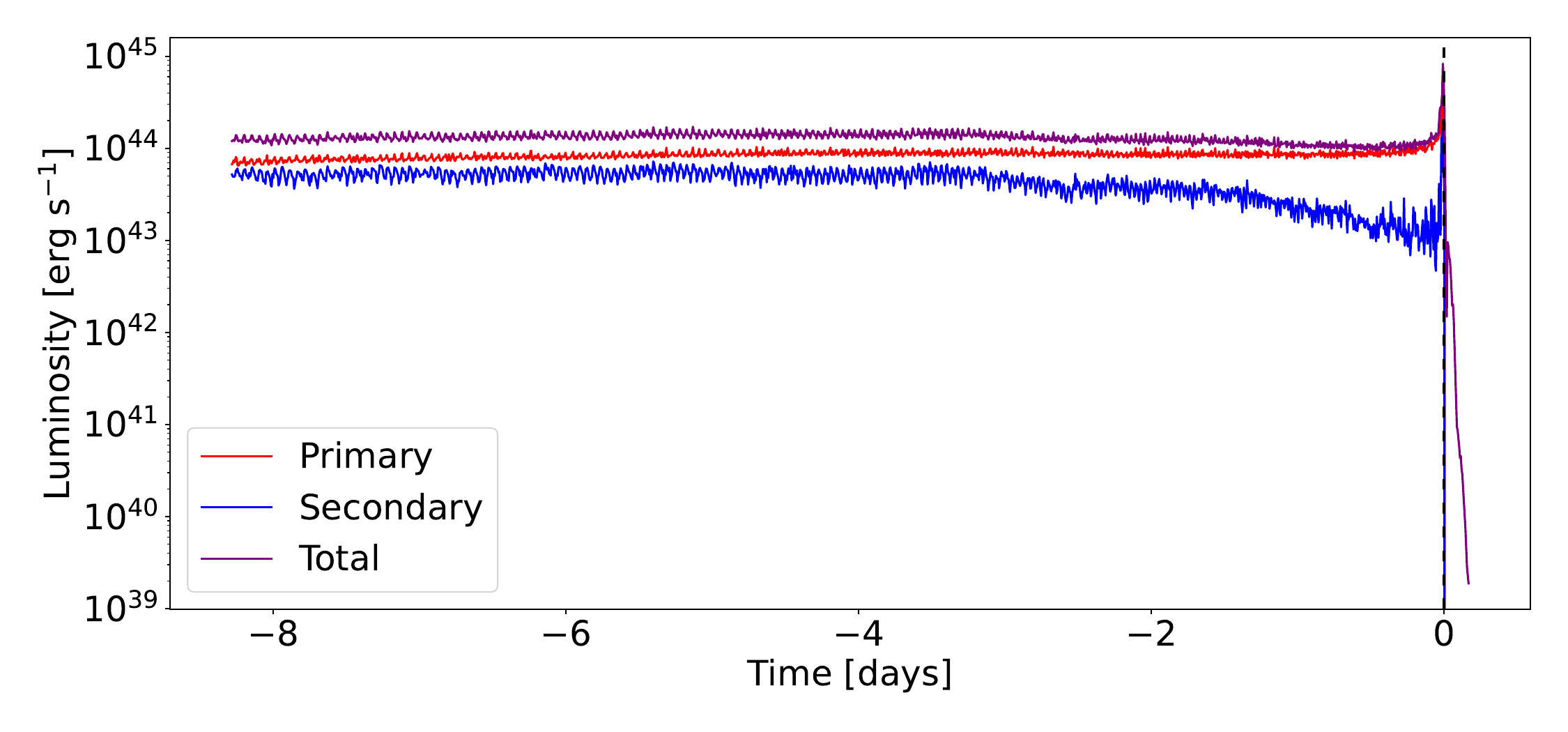}
    \caption{Model $q=0.1$: X-ray light curves for the inspiral to through merger. We see the primary (red) is the dominant contributor to the luminosity throughout merger and at late stages, it outshines the secondary (red) by nearly an order of magnitude.}
    \label{fig:mod4_mds}
\end{figure}

Finally, Figure~\ref{fig:mod4_mds} shows the $q=0.1$ case. Despite the secondary being the dominant accretor early into inspiral, we see now, regardless, the primary is the dominant contributor to the X-ray luminosity. This outshining only becomes more prominent, with the primary being nearly a full order of magnitude brighter than the secondary in the final stages before merger.

Identification of unequal mass binaries using Doppler modulations relies in part on the faster-moving secondary dominating the luminosity. But for sufficiently low $q$, if the primary actually dominates the light curves at late times during inspiral, one may see a disappearance of Doppler modulation instead of an amplification. We test these findings more thoroughly in Appendix~\ref{app-b}.

\subsection{Pre-merger periodicities}
\label{subsec:periods}

In both accretion rates and light curves, there are prominent periodicities on the timescales of both the orbital period of the lump and the orbital period of the binary. However, by looking at the Lomb-Scargle periodograms for each model, we can better understand the dependence on $q$. Specifically, we analyze the periodicity in the X-ray light curves and analytically model the GW inspiral, scaling out the chirp, to extract its frequency for comparison. We begin with the equal-mass case from the \cite{Krauth2023} study.

\begin{figure}
    \centering
    \includegraphics[width=0.45\textwidth]{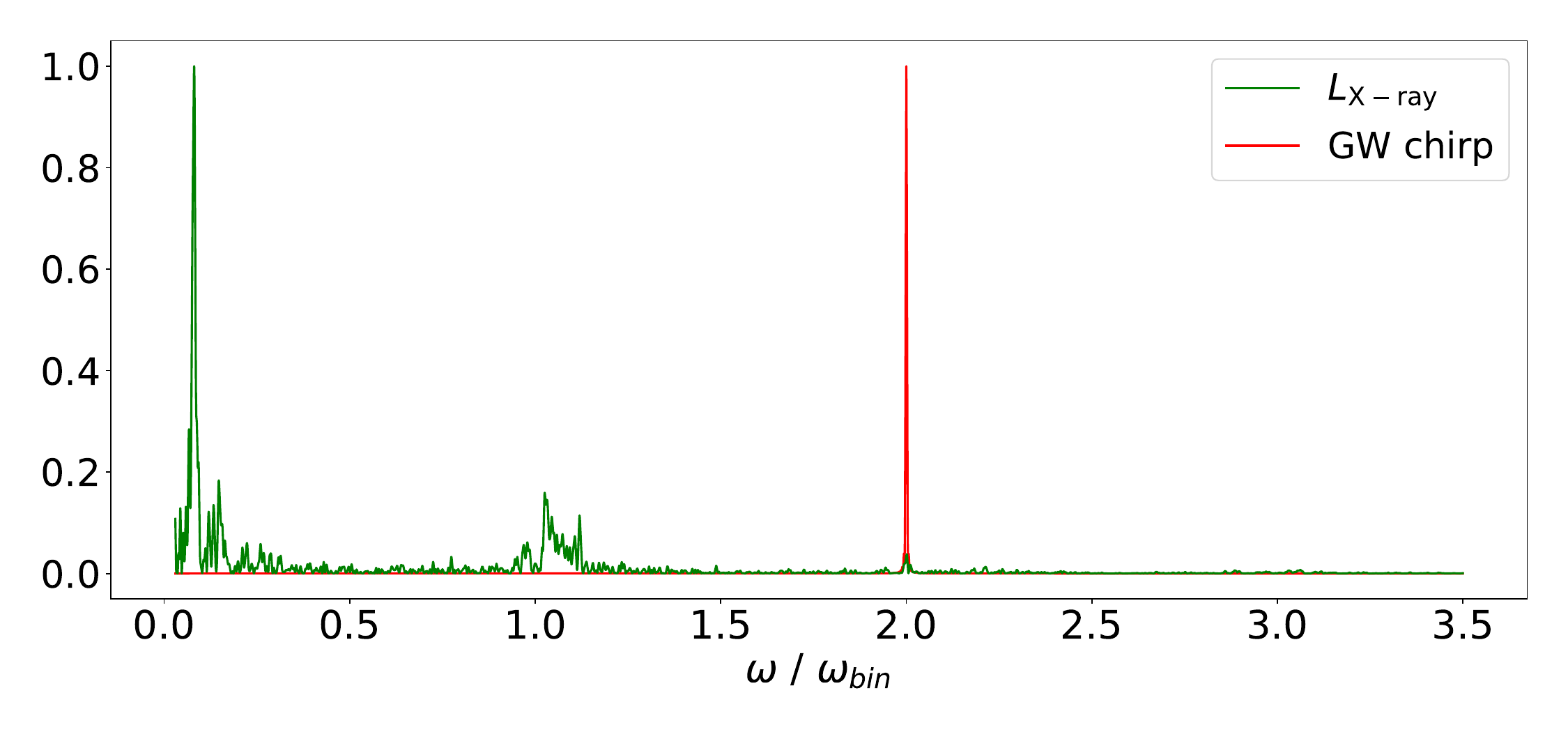}
  \caption{Lomb-Scargle periodograms, with maximum peaks normalized to unity, for the X-ray light curve (green) and GW chirp (red) for the q = 1 fiducial model during the 15 days before merger. 
  Time is measured in units of the instantaneous binary orbital period \protect\cite[adapted from][]{Krauth2023}.}
    \label{fig:pgram_15to0}
\end{figure}

In Figure~\ref{fig:pgram_15to0}, we see the Lomb-Scargle periodograms for both the X-ray light curve (green) and the GW chirp (red) for the last 15 days before merger for the $q=1$ fiducial model of \cite{Krauth2023}. The GW chirp was computed using  Eq.~\ref{eq:peters} for a circular equal-mass binary at $z=1$. Time has been scaled to be in units of the rising instantaneous binary orbital frequency. We see three distinct peaks in the X-ray periodogram. First, the most prominent near $\omega / \omega_{\rm bin} \approx 0.1$, originates from the lump flow pattern propagating along the cavity wall once every 10 binary orbits, which modulates the accretion onto the minidisks and consequently the X-ray light curves originating from them. Next, we see a smaller peak at a value just greater than the binary frequency $\omega / \omega_{\rm bin} = 1$. This primarily originates from mass trading between the minidisks which leads to shock heating of the gas thereby leading to EM flares. This frequency is slightly different than the binary orbital frequency. The likely explanation is that the minidisks become eccentric through a resonant mass-trading instability, which then causes them to precess into each other periodically. For the retrograde precession that we typically find in our simulations, the synodic frequency of mass trading is larger than the binary orbital frequency \citep[an extensive study of this phenomenon is presented in][]{Westernacher-Schneider2023}, although the prograde case is also possible. Finally, there is the smallest peak in line with the frequency of the GWs at $\omega / \omega_{\rm bin} = 2$. This frequency occurs during the late stages of the inspiral and is caused by both minidisks rapidly and equally stripping gas from the CBD and trading it further.

\begin{figure}
    \centering
    \includegraphics[width=0.45\textwidth]{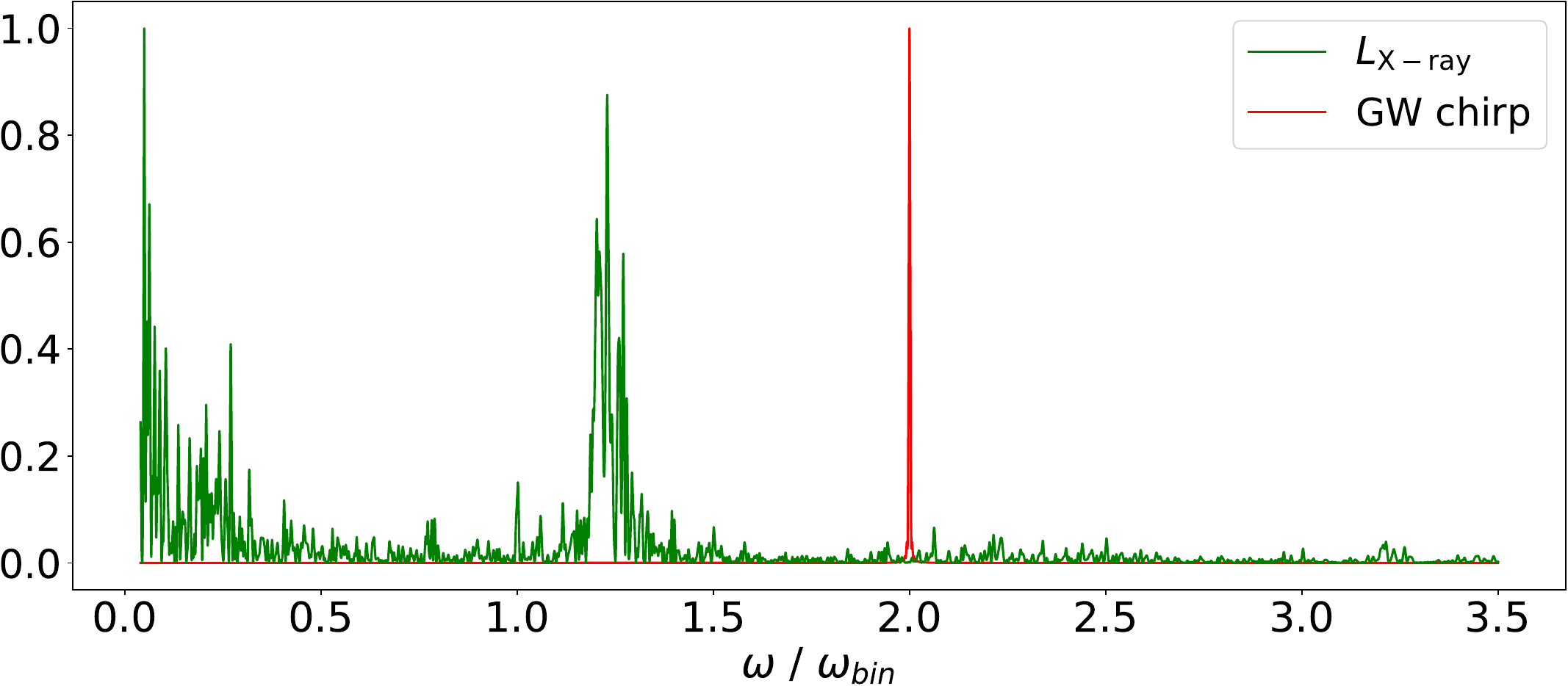}
  \caption{As in Fig.~\ref{fig:pgram_15to0}, but for $q=0.5$. We see a decreased magnitude in the frequency peak associated with the lump's orbit.}
    \label{fig:pgram_q0p5_12to0}
\end{figure}

Next, we move to the periodogram for the full inspiral of the $q=0.5$ model in Figure~\ref{fig:pgram_q0p5_12to0}. Here we begin to see a shift, namely a deterioration of the lump frequency. While there is still a long-term periodicity associated with the lump, we see that it is greatly diminished in that its strength is on equal footing with the peak at the orbital frequency. This corroborates previous findings that the lump feature becomes less distinctive and influential with decreasing $q$ \cite[see, e.g.][]{Dorazio2013,Farris2014,Duffell2020,Siwek2023}. Again we see that the frequency on the orbital timescale is somewhat greater than $\omega / \omega_{\rm bin} = 1$.
 
\begin{figure}
    \centering
    \includegraphics[width=0.45\textwidth]{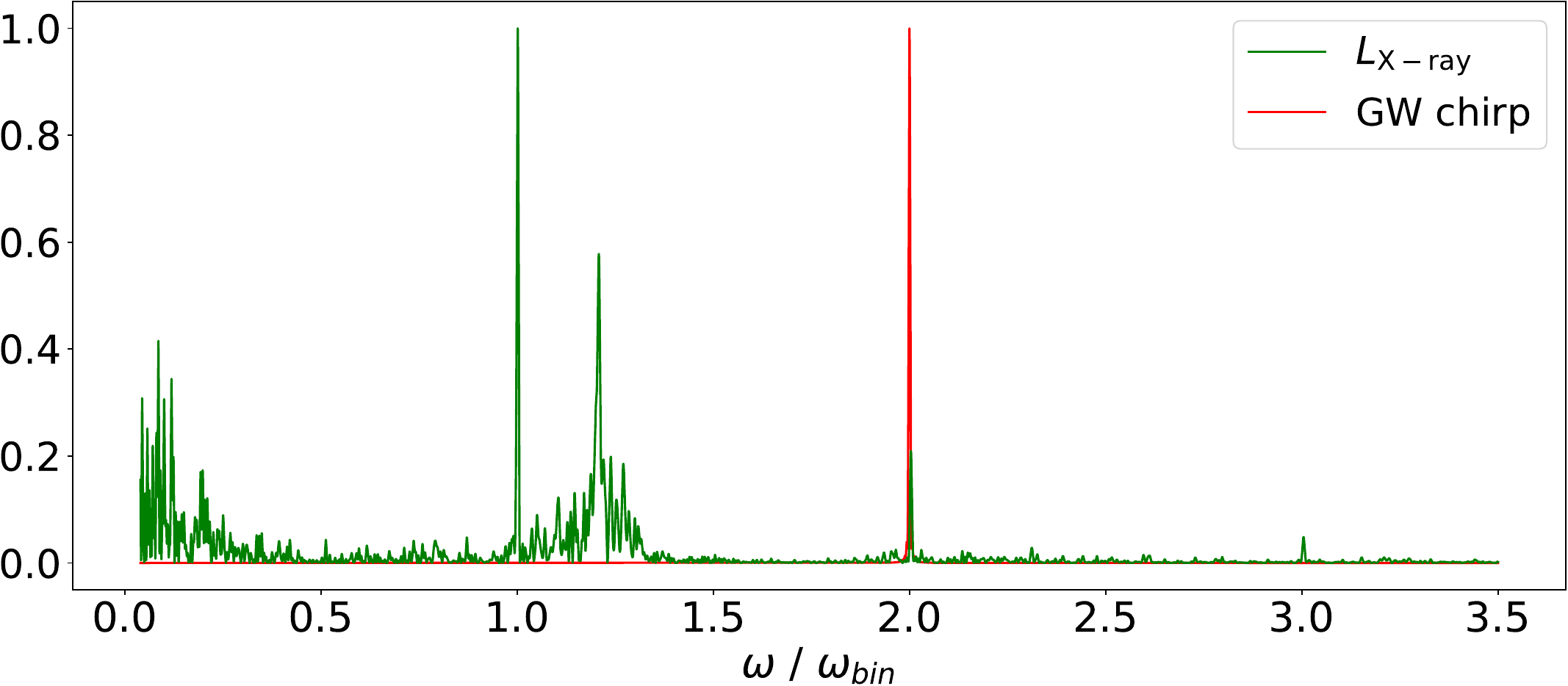}
  \caption{As in Fig.~\ref{fig:pgram_15to0}, but for $q=0.3$. We see further decreasing of the peak associated with the lump's orbit. Additionally, there is a splitting of the frequency associated with the orbital period.}
    \label{fig:pgram_q0p3_10to0}
\end{figure}

Decreasing the mass ratio to $q=0.3$, we see further shifting in Figure~\ref{fig:pgram_q0p3_10to0}. Now, the frequency originating from the lump is sub-dominant, second to the frequency stemming from the orbital frequency. This matches the previously found transitional region between $q\sim 0.5$ to $q\sim 0.3$, in which the lump's presence mitigates significantly \citep{Dorazio2013,Noble2021}. The peak near $\omega / \omega_{\rm bin} = 1$ now has split into two distinct peaks.  While this needs to be further understood, the results are reminiscent in some respect to prior findings. The frequency slightly greater than $\omega / \omega_{\rm bin} = 1$ appears throughout most of the inspiral and seems to originate from the minidisk precession and interaction. However, by inspecting individual portions of the data, we find that the frequency at $\omega / \omega_{\rm bin} = 1$ does not present until late in inspiral ($\sim$10s of orbits before merger). Previous studies have seen near-orbital cadence at larger separations \citep{Westernacher-Schneider2022,Westernacher-Schneider2023}, while others have seen two flares per orbit in late stages \citep{Tang2018,Krauth2023}. We appear to see a similar transition, but now, because inspiral brings the primary further away from the cavity edge, only the secondary tidally interacts with the CBD, and the transition instead is half the frequency of the previous findings (only once per orbit). That is, instead of seeing a shift from just above $\omega / \omega_{\rm bin} = 1$ to $\omega / \omega_{\rm bin} = 2$, we see a shift from just above $\omega / \omega_{\rm bin} = 1$ to $\omega / \omega_{\rm bin} = 1$. If pre-merger monitoring is possible, one may be able to see this shift in dominant period as the merger approaches. Additionally, there are increasingly sub-dominant peaks at $\omega / \omega_{\rm bin} = 2$, and $\omega / \omega_{\rm bin} = 3$, which appear to be overtones of the $\omega / \omega_{\rm bin} = 1$ peak.

\begin{figure}
    \centering
    \includegraphics[width=0.45\textwidth]{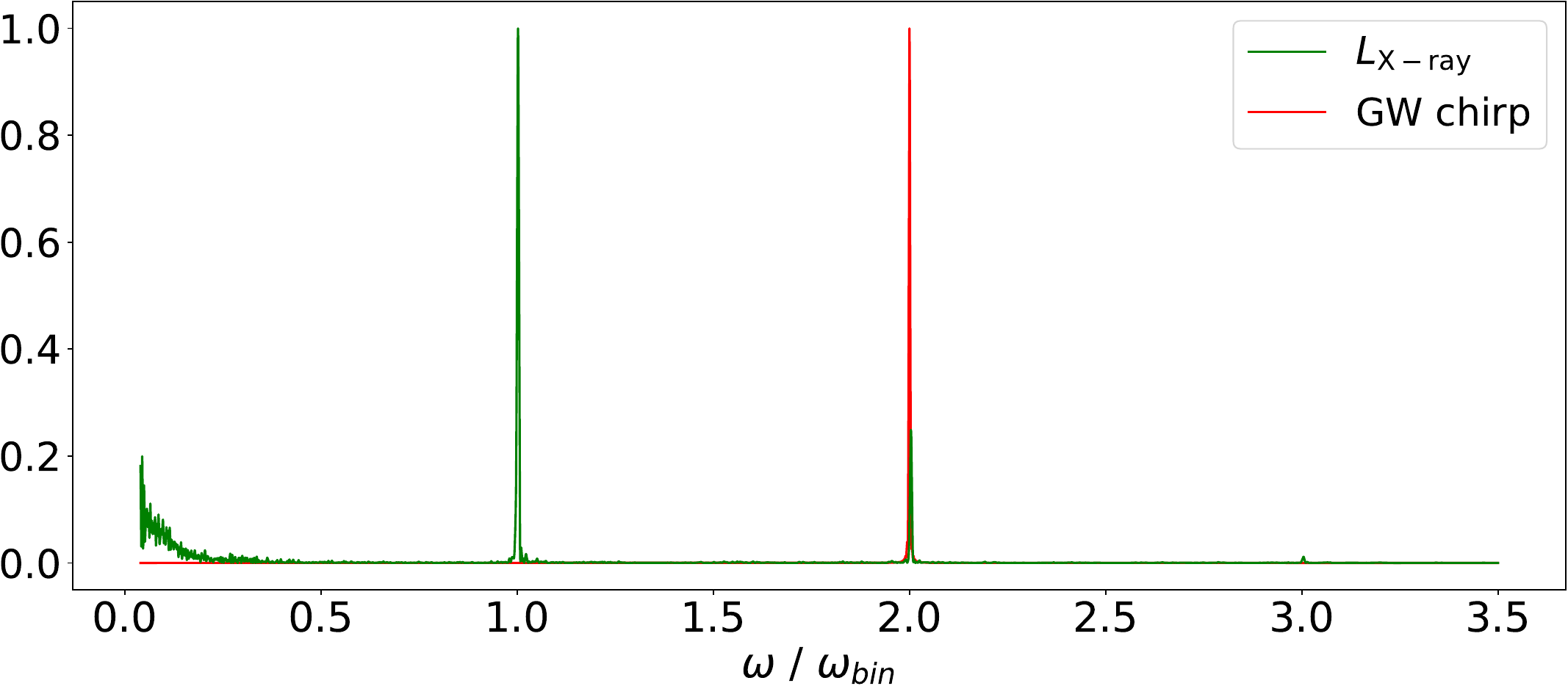}
  \caption{As in Fig.~\ref{fig:pgram_15to0}, but for $q=0.1$. The frequency originating from the lump's orbit is significantly diminished but still present. The frequency coming from the orbital period is now clearly dominant.}
    \label{fig:pgram_q0p1_8to0}
\end{figure}

Finally, we move to the lowest mass ratio of $q=0.1$ of Figure~\ref{fig:pgram_q0p1_8to0}. The frequency associated with the lump's orbit has effectively vanished. There is a strong prominent peak at $\omega / \omega_{\rm bin} = 1$, but again, this presents late into inspiral. Analyzing the inspiral before the $\omega / \omega_{\rm bin} = 1$ peak dominates, there is significantly smaller peak just above this frequency. However, later into inspiral, once again a transition to $\omega / \omega_{\rm bin} = 1$ occurs and it is so prominent that it then dominates the frequency space. There once again appears to be increasingly sub-dominant overtones at $\omega / \omega_{\rm bin} = 2$, and $\omega / \omega_{\rm bin} = 3$.

\section{Conclusions and Discussion}
\label{sec:Con}

We performed 2D hydrodynamical simulations of inspiraling, unequal mass MBHBs from before the nominal decoupling limit through merger. We use an $\alpha$-viscosity disk, directly solve the energy equation with a $\Gamma$-law equation of state for the gas, and incorporate a physically-motivated cooling prescription. We compare our results closely to the fiducial equal-mass study of \cite{Krauth2023} to see the effects of a decreased mass ratio, $q$. We do this for models with mass ratios of $q=0.1, 0.3$ and 0.5. Comparing all of these results, we can draw the following conclusions:

\begin{enumerate}
    
    \item As $q$ decreases, for early times during inspiral, the secondary remains the dominant accretor. However, as time approaches the merger, the roles switch, and the primary becomes the dominant accretor. Our analysis of the data leads us to believe this occurs because of two main reasons. Namely, with decreasing $q$, the tidal influence of the secondary, and its ability to affect gas near the primary, is diminished. Additionally, with lower $q$, the secondary also scoops material from the cavity wall and delivers it directly to the primary, ``force feeding'' the primary. This role-reversal occurs at an earlier stage of the late inspiral with lower $q$.
    
    \item All models regardless of $q$ show a significant drop in accretion rates for both the primary and the secondary BH in the final moments of merger and into post-merger. However, as $q$ decreases a new effect becomes present. The lessened tidal destruction but increased force feeding from the secondary, causes a rapid rise in the accretion rate of the primary, and a brief but sharp spike in accretion of the secondary just before the drop in accretion occurs.

    \item All models regardless of $q$ show a corresponding several order of magnitude drop in the thermal X-ray luminosity from pre- to post-merger (as well as a smaller drop in the UV luminosity). While the magnitude of the drop remains comparable in all models, as $q$ is lowered, the timing of the drop changes. In the equal-mass case the drop occurs several hours beforehand, whereas with all lower $q$ models, the drop occurs at 10s of minutes before merger. This drop occurs because the minidisks, which are responsible for nearly all of the X-ray luminosity, are increasingly tidally truncated with inspiral leading to their destruction and the consequential drop in X-ray luminosity.

    \item At low enough $q$ a potential new X-ray signature appears. On the order of hours before merger, the decreased tidal destruction and increased force feeding from the secondary lead to an excitation of the gas in the primary minidisk, and a sharp spike in the X-ray luminosity before the inevitable plummet. The secondary minidisk, while having sub-dominant luminosity in the late inspiral, also flares around the same time as the primary. This effect becomes more prominent at low $q$. We test these findings against numerical effects by altering our burn-in period, grid and sink refinement, our sink prescription, and our sink rates in Appendix~\ref{app-a}. Regardless of each of these changes, we still see the pre-merger flaring. However, its amplitude can vary. It will require further testing to determine whether the pre-merger flare would be distinguishable from noise, and whether that flare is useful for binary identification or parameter estimation.
    
    \item One important conclusion from these findings is that if the fuller, force-fed primary minidisk in lower-$q$ systems dominates the light curve just before merger, this will reduce the Doppler modulation, because the velocities of the primary and secondary scale inversely with mass $v_1\sim v_2 M_2/M_1$. Previously, it has been assumed that the secondary dominates \citep[e.g][]{Haiman2017} when predicting the Doppler modulation amplitude. If sufficiently continuous pre-merger monitoring of the source is possible, this may reduce the ability to find Doppler modulations in the last moments before merger. However, it may also lead to another signature, namely the disappearance of Doppler modulations when the transition of the dominant minidisk occurs. Furthermore, if the primary dominates the X-ray light curve for the entire inspiral (as our $q=0.1$ case seems to suggest), then perhaps the ability to use Doppler modulation for identification of sufficiently low-$q$ binaries may be compromised entirely. Further study of the Doppler modulations from such systems will be needed.
    
    \item Lomb-Scargle periodograms of the X-ray light curves show two primary frequencies. One originates from the overdense lump, and the other from the orbital motion of the binary. The former dominates at higher $q$, and the latter dominates at lower $q$, with the transition occurring somewhere near $q=0.3$.

    \item The Lomb-Scargle periodograms are time-dependent. At high $q$ a late inspiral frequency of $\omega / \omega_{\rm bin} = 2$ presents, while at low $q$, we instead see a frequency of $\omega / \omega_{\rm bin} = 1$. We attribute this to the primary's decreased interaction with the CBD at low $q$ during late inspiral.
    
\end{enumerate}

One of the primary goals of these signatures would be to couple the unique EM signature with the GW observations of LISA to aid in sky localization before merger occurs. \cite{Krauth2023} summarizes the prospect of this for the upcoming X-ray telescopes of the Advanced Telescope for High-energy Astrophysics \citep[Athena;][]{Nandra2013} and for the proposed Lynx mission \citep{Lynx2018}. The primary takeaway is that either of these telescopes should be able to search LISA's entire error box 10 hours prior to merger out to $z=0.5$ for hard X-rays in the 2-10 keV band, or out to $z=1$ for soft X-rays in the 0.5-2 keV band in the case of Athena, and out to $z=1$, regardless of X-ray hardness, for Lynx \citep{Lops2023}. \cite{Krauth2023} points out that as little as two data points, one a bit earlier on and one occurring during the X-ray drop, could be used to identify the correct host galaxy before merger for an equal-mass binary. Knowledge from LISA about when the binary will merge would aid in selecting observational times that are likely to reveal the X-ray drop. This could reduce the burden of the high observational cadence needed when looking for extended pre-merger periodicity in a limited time window - an effect that also increasingly diminishes for low-$q$ binaries.

For the unequal mass binaries studied here, the tell-tale thermal X-ray drop occurs, but the window to obtain the second data point is more limited. If, however, a sharp increase in the X-ray luminosity before the drop occurs for lower mass ratio systems, we can re-extend that window to hours instead of 10s of minutes before the merger. By searching for the X-ray increase before the drop, we may still be able to identify the host galaxy before merger with as few as two data points. Additionally, if we have more data points, for both the increase and the drop, we may also be able to use this information to estimate the mass ratio through EM means. For sources with a sufficient signal-to-noise ratio (SNR), LISA can also obtain $q$ through GW measurements. If comparison of these two values differ, it may allow us to learn more about the circumbinary gas dynamics. Regardless, the host galaxy becoming X-ray dark and staying that way for a significant time after merger should enable confident post-merger identification of the correct host galaxy.

As the potential caveats to these findings are already itemized in \cite{Krauth2023} we will not restate all of them here, but refocus on a few of them which could potentially affect these unequal-mass findings more specifically. Some of the important differences from \cite{Franchini2024} are the implementation of PN corrections and setting the sink radius at the Innermost Stable Circular Orbit (ISCO) instead of the BH horizon. These effects can cause the drop to occur sooner during inspiral as the gas nearest the BH is lessened. Regardless, they still find a drop in thermal X-ray luminosity occurring most significantly just hours before merger. Inclusion of a pseudo-Newtonian potential \citep[e.g.][]{Paczynski1980} could still have important effects for the unequal mass case. If the minidisks are dismantled sooner during inspiral, perhaps the spike in luminosity before the drop is mitigated, or perhaps the drop occurs sooner, implying a longer time window to detect it. Additionally, either radiative feedback \citep[e.g.][]{Delvalle2018} that could also help destroy the minidisks from the inside, or BH spins~\citep[e.g.][]{Paschalidis2021,Combi2022} which can alter the composition of the minidisks, could have similar effects. Further exploration of the parameter space is left to future research.

\section*{Acknowledgements}

We acknowledge support from the NWA Roadmap grant ``GW LISA/ET: Shivers from the Deep Universe: a National Infrastructure for Gravitational Wave Research'' (LK), NSF grant AST-2006176 (ZH), and NASA ATP grant 80NSSC22K0822 (AM and ZH). JD is supported by NASA through the NASA Hubble Fellowship grant HST-HF2-51552.001A, awarded by the Space Telescope Science Institute, which is operated by the Association of Universities for Research in Astronomy, Incorporated, under NASA contract NAS5-26555. This research was supported in part by the National Science Foundation under Grant No. NSF PHY-1748958. This research has made use of NASA's Astrophysics Data System.
{\it Software:} {\tt python} \citep{travis2007,jarrod2011}, {\tt scipy} \citep{jones2001}, {\tt numpy} \citep{walt2011}, and {\tt matplotlib} \citep{hunter2007}.

\section*{Data Availability}
The data underlying this article will be shared on reasonable request to the corresponding author.


\bibliographystyle{mnras}

\appendix

\section{Sensitivity Tests}
\label{app-a}

Here we test the sensitivity of our results to the grid refinement as well as the sink refinement, prescription, and rates in our simulations. In our main text runs, the sink prescriptions and softening length are also evolved during the steps in which we increase of resolution. Starting with equal sink radii and softening lengths for each mass, we increase (decrease) the primary (secondary) sink radius and softening length each time we increase the resolution. The first step is halfway to the final value, and the final step is set to $r_{\rm s,bh}$ for each black hole. We test this for the $q=0.3$ unequal mass case, in which case we first see both the new transient of the flare appear and in which the primary becomes the dominant accretor in the final stages of inspiral.

\begin{figure*}
    \centering
    \includegraphics[width=0.9\textwidth]{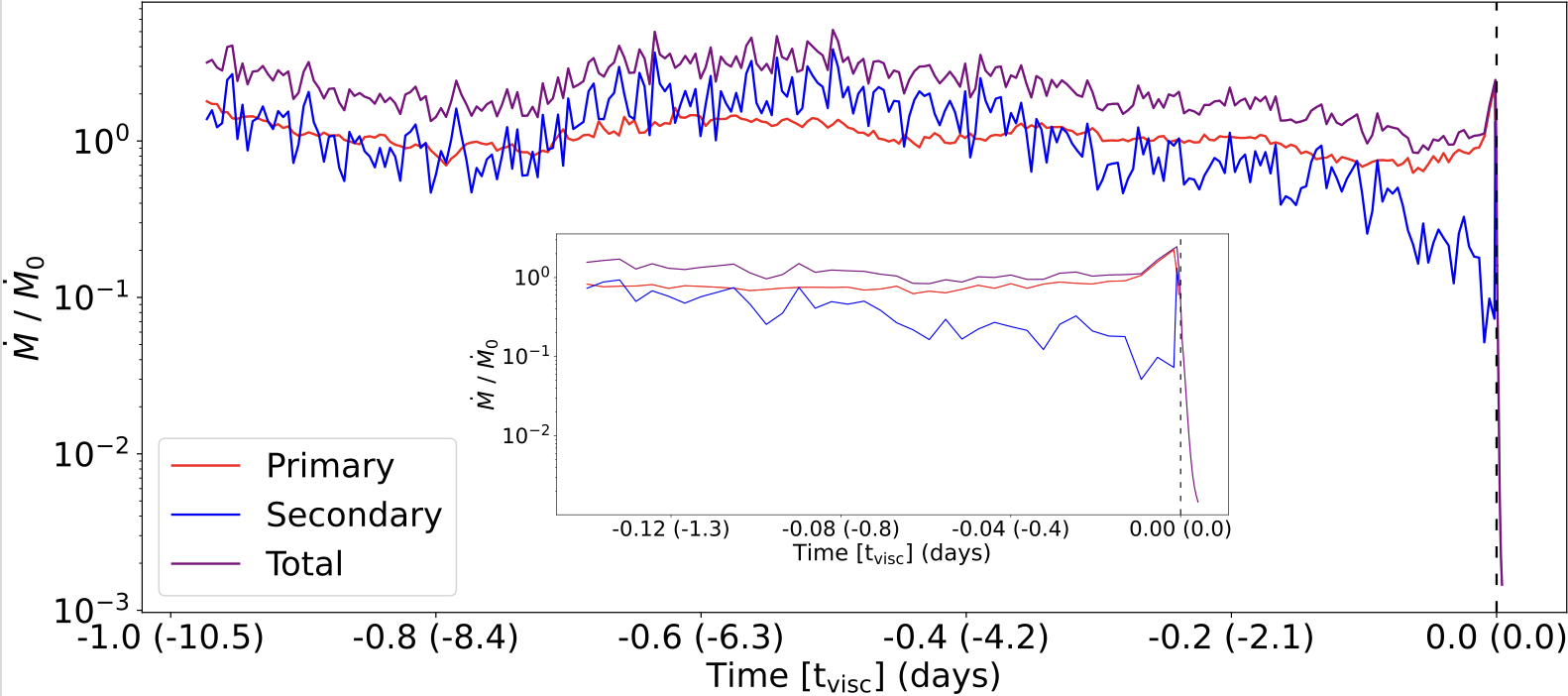}
    \caption{Accretion rates for $q=0.3$ model while using the acceleration-free sink model, without burn-in or refinement. Comparing to the previous $q=0.3$ accretion rates in Fig.~\ref{fig:mod5_acc} we see many similarities. Once again as we approach merger that dominant accretor becomes the primary. Once again, we see a spike in the accretion rates of the primary and the secondary shortly before merger. Again though, we see the accretion for both BHs begins dropping before merger occurs. Accretion again plummets several orders of magnitude into post-merger as in our previous $q=0.3$ model.}
    \label{fig:af_mod5_acc}
\end{figure*}

Instead of running the simulation for a long burn-in time period to allow for relaxation of the CBD, we instead run a simplified model, initializing inspiral at the start of the simulation. In our first test, instead of refining the grid resolution and sink size in time, we run the entire simulation with the final resolution and sink sizes from the start of the simulation. Additionally, we use an acceleration-free sink prescription in which the sinks remove both mass and angular momentum, instead of the torque-free model in our main text runs, in which the sink does not absorb gas angular momentum as defined in its own instantaneous rest frame. In Figure~\ref{fig:af_mod5_acc}, we see the accretion rates. The primary accretor at early times is less clear, however, this could simply be due to the lack of initial burn-in phase before initializing inspiral. We do however see that the primary becomes the dominant accretor around the same time ($\sim0.4$ viscous times remaining). More importantly, looking to the inset plot, we see both the characteristic flare and drop in the accretion rates of the primary and secondary leading to merger, consistent with the $q=0.3$ run. However, the spike is less severe, by a factor of $\sim2-3$ less than the previous case. While this indicates the magnitude of the features may depend on the parameters such as the burn-in period, grid refinement, sink refinement, or sink prescription, the qualitative signatures of flaring and dropping in the accretion merger merger do not.

\begin{figure*}
    \centering
    \includegraphics[width=0.9\textwidth]{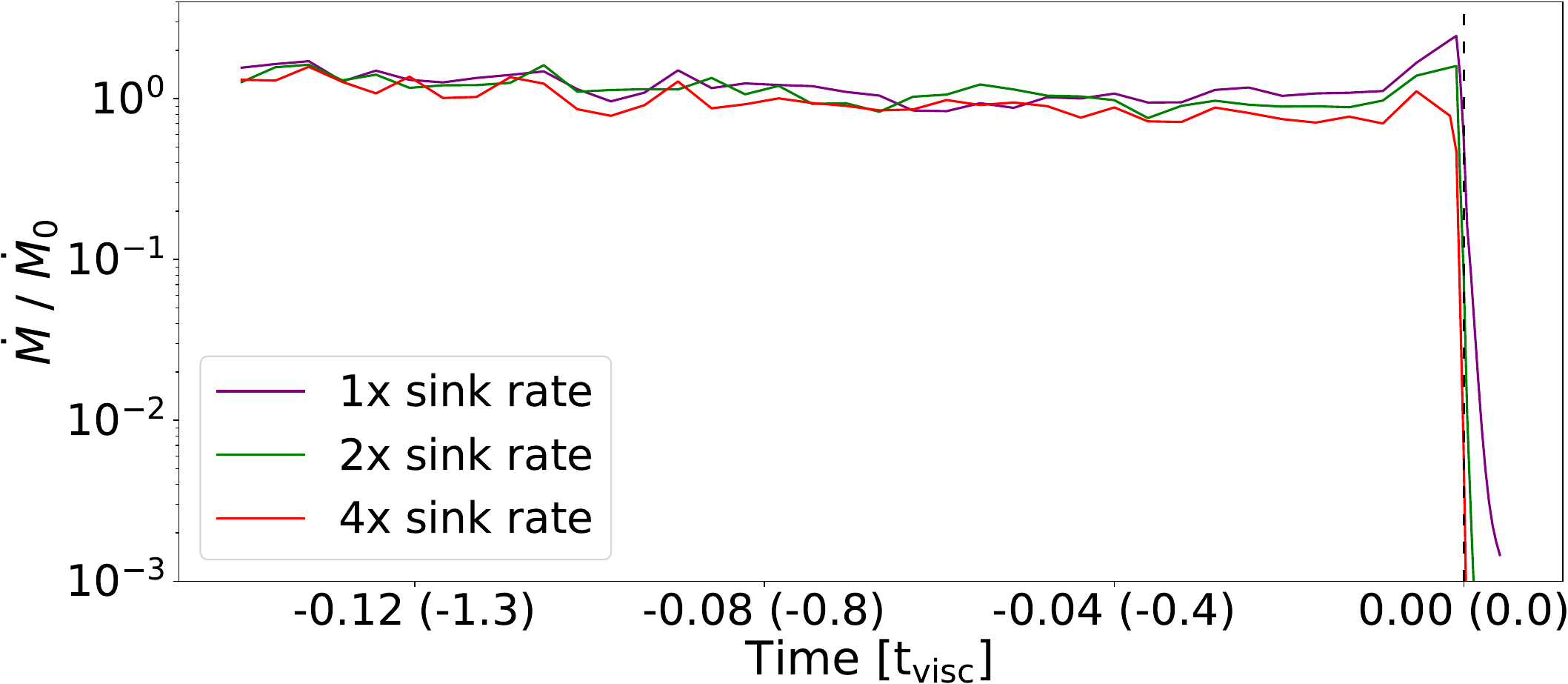}
    \caption{Total accretion rates for $q=0.3$ acceleration-free sink model, without burn-in or refinement, varying the sink rate to 2 and 4 times our initial sink rate. While we still see a spike in the accretion rate shortly before merger, higher (perhaps unphysical) sink rates do diminish the spike by a factor of $\sim2$. Regardless, we again see the accretion begins dropping before merger occurs. Accretion again plummets several orders of magnitude into post-merger as in our previous $q=0.3$ models.}
    \label{fig:af_compare}
\end{figure*}

Additionally, we check to see how the results depend on sink rates, seen in Figure~\ref{fig:af_compare}. The former plot has a sink rate chosen to be several times larger than the viscous rate at the sink edge and should therefore be more than sufficient to remove material. However, here we increase these rates by a factor of 2 and 4, to see how this affects our findings. In increasing the sink rates, we do see an effect on the accretion rates just before merger during the flare. It seems when the rates are significantly high, as in the 4$\times$ case, there is a factor of $\sim$2 decrease in the amplitude of the flare. While this does seem to affect results modestly, the characteristic flaring and drop still occur and it should be noted that raising the rates this significantly may be unphysical.

\section{Brighter Minidisk Tests}
\label{app-b}

\begin{figure}
    \centering
    \includegraphics[width=0.45\textwidth]{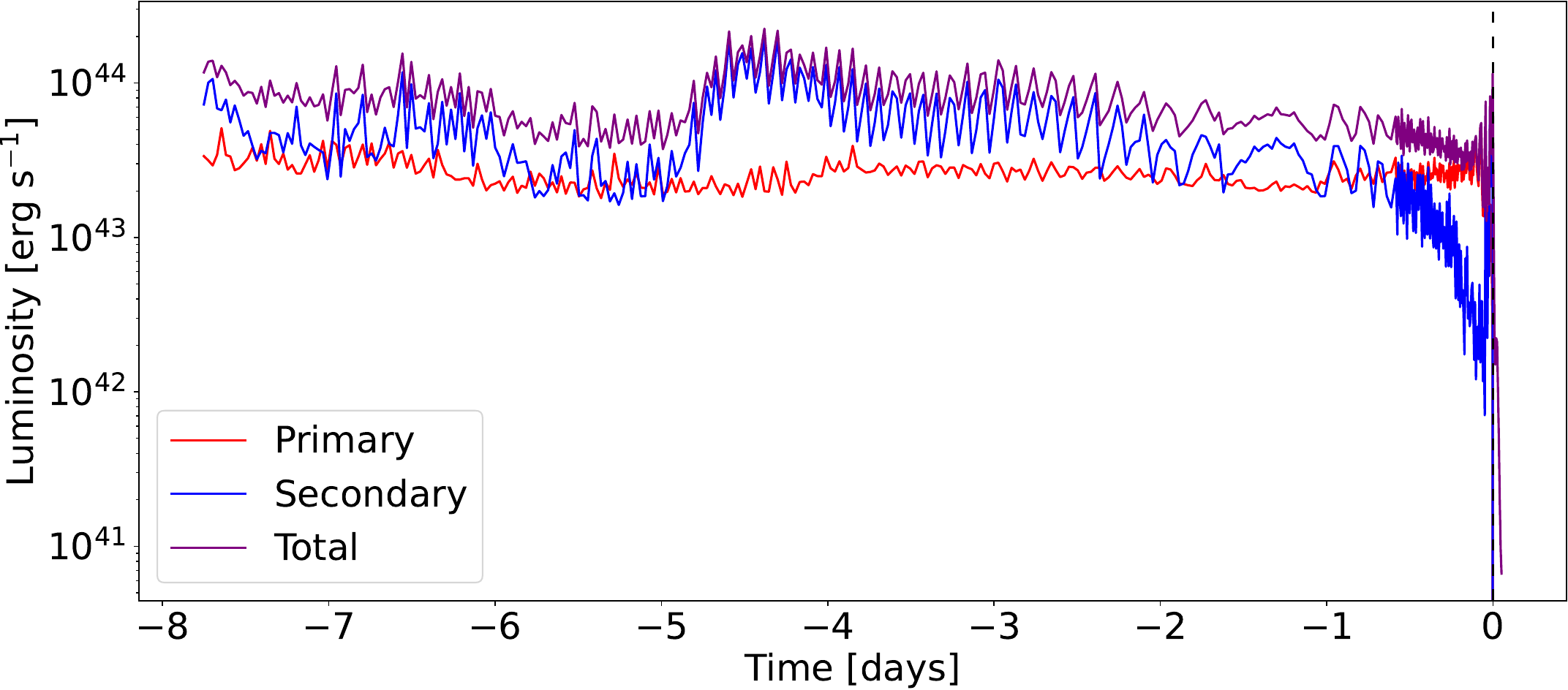}
    \caption{Model $q=0.1$: X-ray light curves for the inspiral to through merger for acceleration-free sinks. We see the primary (red) contribution to the overall X-ray luminosity is less than the secondary (blue) contribution until the final moments before merger.}
    \label{fig:noburn_af_mod4_mds}
\end{figure}

\begin{figure}
    \centering
    \includegraphics[width=0.45\textwidth]{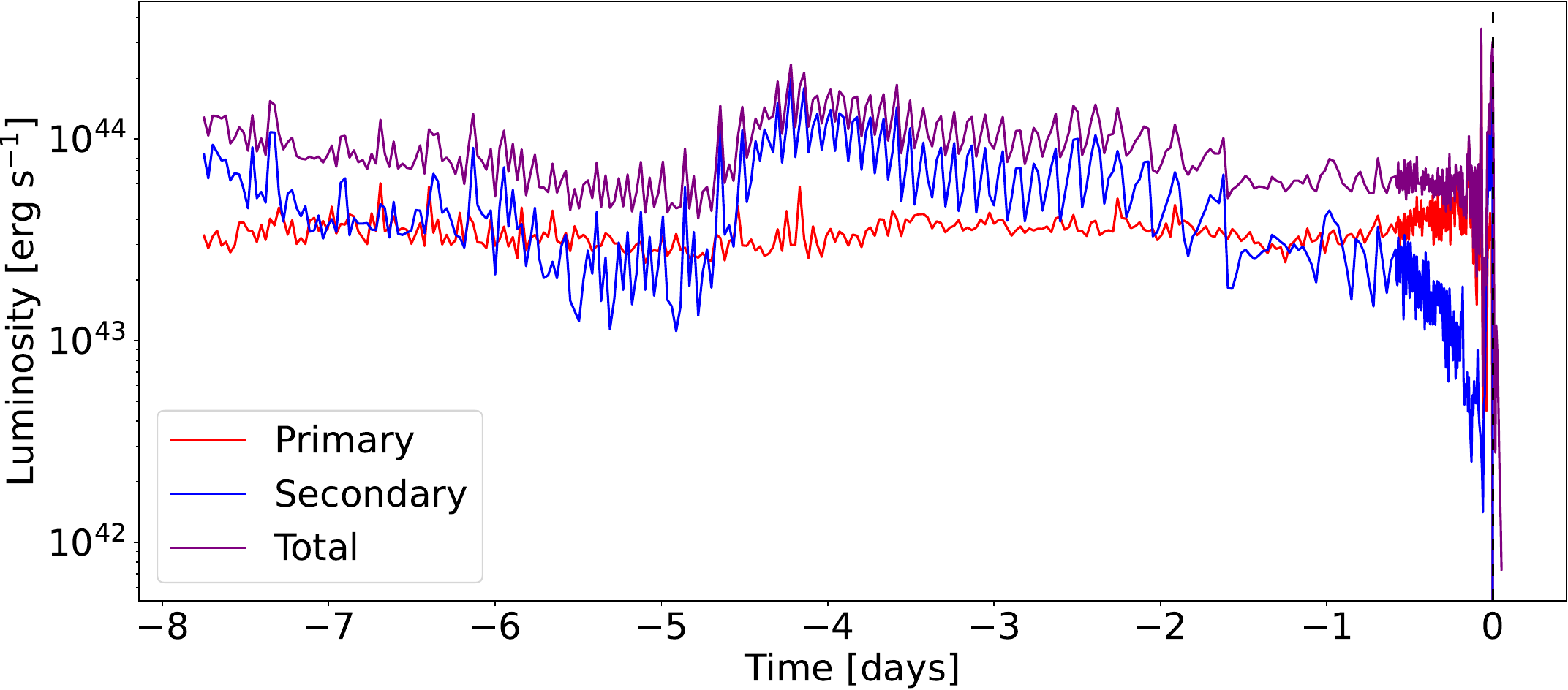}
    \caption{Model $q=0.1$: X-ray light curves for the inspiral to through merger for torque-free sinks. We see the primary (red) contribution to the overall X-ray luminosity is less than the secondary (blue) contribution until the final moments before merger.}
    \label{fig:noburn_tf_mod4_mds}
\end{figure}

As finding which component shines brightest can be important for Doppler modulation signatures, here we alter our simulations to investigate the robustness of this finding. Once again, we run without burn-in, as well as without grid and sink refinement. We test both acceleration-free and torque-free sink models for the $q=0.1$ case, in which the primary dominates the luminosity most significantly. We see in Figures \ref{fig:noburn_af_mod4_mds} and \ref{fig:noburn_tf_mod4_mds} in both the acceleration-free and the torque-free sink models the primary outshines the secondary during the late inspiral. However, we see in both models when compared to our fiducial $q=0.1$ model in Figure \ref{fig:mod4_mds}, that the secondary tends to more commonly be brighter earlier into inspiral. These tests suggest that the brighter primary found during the entire q=0.1 inspiral is not achieved without a burn-in period, and is unlikely to be dependent on sink prescription. This highlights the importance of allowing the gas to sufficiently relax before initiating inspiral. Further testing in this area is left to future work.

\label{lastpage}
\end{document}